\documentclass[pdflatex,sn-mathphys-num]{sn-jnl}

\usepackage{graphicx}
\usepackage{amsmath,amssymb,amsfonts}
\usepackage{xcolor}
\usepackage{booktabs}
\usepackage{float}
\usepackage{array}
\usepackage{enumitem}
\usepackage[most]{tcolorbox}
\usepackage{tabularx}
\hypersetup{hypertexnames=false}

\floatstyle{ruled}
\newfloat{algorithm}{htbp}{loa}
\floatname{algorithm}{Algorithm}

\begin{document}

\title[Class-level LLM-generated test suites in Python]{Evaluating the effectiveness of class-level LLM-generated test suites in Python}


\author*[1,2]{\fnm{Bilal} \sur{Al-Ahmad}}\email{balahmad@aus.edu}

\author[3]{\fnm{M.} \sur{Harshvardhan}}\email{harshvardhan@aus.edu}

\author[1]{\fnm{Khaled} \sur{El-Fakih}}\email{kelfakih@aus.edu}

\author[4]{\fnm{Anas} \sur{AlSobeh}}\email{anas.alsobeh@uvu.edu}

\affil[1]{\orgdiv{Department of Computer Science and Engineering},
\orgname{American University of Sharjah},
\orgaddress{\city{Sharjah}, \country{United Arab Emirates}}}

\affil[2]{\orgdiv{Department of Computer Information Systems},
\orgname{The University of Jordan},
\country{Jordan}}

\affil[3]{\orgdiv{Information Systems and Analytics},
\orgname{American University of Sharjah},
\orgaddress{\city{Sharjah}, \country{United Arab Emirates}}}

\affil[4]{\orgdiv{Information Systems and Technology \- Applied AI},
\orgname{Utah Valley University},
\orgaddress{\city{Orem}, \state{Utah}, \country{United States}}}

\abstract{
	\textbf{Context} Large language models (LLMs) can generate unit tests quickly, but high structural coverage does not establish that those tests execute reliably or detect faults. Existing evidence often treats coverage as the principal outcome and rarely compares prompt strategies and models through mutation testing at class level. \\
	\textbf{Objective} This study examines how prompt strategy and model choice shape the executability, structural coverage, fault-detection effectiveness, and structural quality of LLM-generated Python test suites relative to human-written suites. \\
	\textbf{Method} We evaluate multiple prompt strategies across a diverse set of current LLM configurations on the ClassEval benchmark. The evaluation combines execution outcomes, line and branch coverage, Cosmic Ray mutation scores, and structural quality indicators. Primary analyses treat successful execution as a prerequisite; paired comparisons use only classes shared by the relevant executable subsets. \\
	\textbf{Results} Structural coverage is consistently near its ceiling and offers little discrimination among configurations. Executability varies substantially. The proposed prompt performs strongly for mutation score, but no prompt dominates across models. Model choice explains more variation than prompt choice, and their interaction shows that prompt effectiveness depends on the selected model. Human and LLM suites are evaluated on unequal executable subsets, so their relative mutation scores do not establish superiority. \\
	\textbf{Conclusions} Reliable assessment of LLM-generated tests should treat executability as a gate and combine coverage with mutation testing and structural quality indicators. In practice, model selection should precede prompt tuning.}


\maketitle

\section{Introduction}\label{sec:intro}

\subsection{Problem Statement}\label{sec:problem}

Large language models (LLMs) now support software-development tasks ranging from code completion and documentation to bug detection and repair. Automated test generation is a consequential application because writing and maintaining tests requires substantial developer effort. Tools such as GitHub Copilot and Amazon CodeWhisperer have accelerated industrial adoption, while research systems such as ChatUniTest and CoverUp target test generation directly \citep{chen2024chatunitest,pizzorno2025coverup}. LLMs can produce syntactically valid test files within seconds, but speed alone says little about test quality.

Evaluation remains dominated by structural metrics such as line and branch coverage \citep{wang2025projecttest,wang2025testeval}. Coverage records execution; it does not establish fault detection. A suite can reach 100\% line coverage while making no meaningful assertion about program behavior. This distinction is well established in software-testing research~\citep{inozemtseva2014coverage}, but it is particularly important for LLM-generated tests. Language models can exercise many paths without producing assertions that distinguish correct behavior from faulty behavior. High coverage therefore leaves the central question unanswered: do the generated tests find bugs?

\subsection{Research Motivation}\label{sec:motivation}

Mutation testing addresses this gap directly. By introducing small, systematic faults (mutants) into the source code and measuring whether a test suite detects them, mutation scores provide a rigorous proxy for fault detection capability~\cite{jia2011analysis}. Several recent studies have begun incorporating mutation analysis into LLMs test evaluation~\cite{guilherme2023initial, nafis2025llm}, but these investigations typically examine a single prompt strategy with one or two models. The interaction between prompt design and mutation effectiveness remains unexplored.

A parallel limitation concerns prompt engineering itself. The prompts used to elicit test suites from LLMs vary enormously, from minimal instructions of 18 lines~\cite{ouedraogo2024large} to elaborate chain-of-thought frameworks spanning 77 lines with explicit quality criteria and structural guidance~\cite{chudic2026automated, wang2025testeval}. Practitioners face a practical question with no empirical answer: does investing in sophisticated prompt design yield better tests? Despite the centrality of this question, no controlled comparison exists that holds the model, benchmark, and evaluation methodology constant while varying only the prompt.

This study addresses both gaps simultaneously. We evaluate four prompt strategies of varying complexity across {ten} LLMs configurations on 100 Python classes from the ClassEval benchmark~\cite{du2024classeval}, using a multi-dimensional evaluation framework that incorporates Cosmic Ray mutation scores alongside structural coverage, Executable Test Percentage, and test quality metrics. The four prompts span a range of design philosophies: our proposed prompt (77 lines) incorporates an explicit analysis phase, mutation-resistance instructions, and detailed quality standards; Wang et al.'s TestEval prompt~\cite{wang2025testeval} (30 lines) provides structured task descriptions with coverage targets; Chudic and \c{C}al{\i}kl{\i}'s prompt~\cite{chudic2026automated} (29 lines) adopts a few-shot style with example references; and Ouedraogo et al.'s prompt~\cite{ouedraogo2024large} (18 lines) uses minimal zero-shot instructions. {The ten LLMs configurations span four providers and multiple reasoning levels}: GPT-5.2 with no reasoning and medium reasoning, Gemini 3 Flash with low and high reasoning, {DeepSeek in both standard (Chat) and reasoning (Reasoner) modes, and Anthropic Claude (Opus 4.6 and Sonnet 4.6), each without and with extended thinking}.

The results challenge conventional assumptions about prompt engineering for test generation. Structural coverage is {at or near} ceiling for every (model, prompt) cell, confirming that coverage alone cannot discriminate between prompt strategies. No single prompt dominates on fault detection. Our proposed prompt is the modal best choice, winning on {seven of ten} model configurations; Ouedraogo et al.\ wins on {two (GPT-5.2 with medium reasoning and Claude Sonnet without extended thinking)}; Chudic \& \c{C}al{\i}kl{\i} wins on one (DeepSeek Chat). The human baseline kills 88.3\% of mutants on the 82 classes where its suites run, while the {ten} LLMs configurations cluster between 90.2\% and 96.2\%; LLMs kill mutants at similar or slightly higher rates than humans on the same classes, while humans remain more reliably executable. {Model selection exerts an effect on mutation scores roughly 3.9 times larger than prompt choice ($\eta^2 = 0.0216$ vs.~$\eta^2 = 0.0056$). After Holm-Bonferroni correction, 81 of 703 eligible pairwise differences survive, and although the prompt main effect in a two-way ANOVA now reaches significance ($p = 0.007$), it remains roughly a quarter the size of the model effect.} These findings suggest that the LLM's intrinsic capabilities constrain test quality far more than the instructions it receives.

\subsection{Research Questions}\label{sec:rqs}

We investigate three research questions that isolate the contribution of prompt design from that of model selection:

\begin{itemize}
    \item \textbf{RQ1:} Does prompt strategy significantly affect structural coverage (line and branch) of LLM-generated tests?
    \item \textbf{RQ2:} Does prompt strategy significantly affect fault detection capability, as measured by mutation score?
    \item \textbf{RQ3:} What is the relative importance of model selection versus prompt design for test quality?
\end{itemize}

\noindent We note upfront that we find no statistically significant differences between prompt strategies on line and branch coverage when test suites execute successfully. This negative result is itself informative: it demonstrates that coverage cannot discriminate between prompt strategies that differ substantially on fault detection effectiveness, validating our decision to adopt a multi-dimensional evaluation framework.

\subsection{Contributions}\label{sec:contributions}

{This paper makes the following contributions.}
\begin{itemize}
    \item \textbf{A factorial prompt-by-model study.} We cross four prompt strategies with {ten} LLMs configurations on 100 class-level Python programs, yielding {4,000} generated test suites plus 100 human baselines, and we hold benchmark, tooling, and evaluation protocol constant so that prompt and model effects can be separated.
    \item \textbf{A multi-dimensional, executability-gated evaluation framework.} We combine Executable Test Percentage, line and branch coverage, Cosmic Ray mutation score, and structural quality metrics, and we condition every downstream comparison on the classes where the suite actually executed, rather than scoring non-executing suites as zero.
    \item \textbf{The first systematic application of mutation testing across multiple prompts and multiple LLMs on class-level Python code,} together with the empirical finding that model selection dominates prompt design ({$\eta^2_{\mathrm{model}}/\eta^2_{\mathrm{prompt}}=3.9$}) and that executability, not mutation score, is the binding constraint that most distinguishes configurations.
    \item \textbf{A proposed zero-shot prompt with embedded chain-of-thought} for class-level test generation, evaluated against three literature baselines, and released together with the full evaluation pipeline.
\end{itemize}

\subsection{Paper Organization}\label{sec:organization}

The remainder of this paper is organized as follows. 
\autoref{sec:background} introduces the LLM configurations, test-generation prompts, and evaluation metrics. 
\autoref{sec:relatedwork} reviews LLM-based test generation and prompt engineering before identifying the research gap. 
\autoref{sec:methodology} describes context extraction, prompt design, benchmarking, output post-processing, and the end-to-end pipeline. 
\autoref{sec:results} reports the experimental evaluation and statistical analyses. 
\autoref{sec:discussion} answers the research questions, relates the findings to prior work, and examines implications and validity threats. 
\autoref{sec:conclusion} summarizes the study and outlines future work.

\section{Preliminaries}\label{sec:background}

This section defines the concepts used throughout the study: the LLMs and reasoning (``thinking'') levels, the role of prompts in test generation, and the metrics used to evaluate generated suites. 

\subsection{Large Language Models and Levels of Thinking}\label{sec:llm-thinking}

Modern code-capable LLMs are autoregressive models trained on large corpora of source code and natural language. Given a prompt containing a unit under test, they emit a candidate test file token by token. Providers now expose \emph{reasoning} or \emph{thinking} controls that allocate additional internal computation before the model returns its answer. More reasoning can improve the structure and completeness of generated code, but it also increases latency and cost and does not guarantee an executable suite. We evaluate ten configurations from four providers: OpenAI GPT-5.2 without reasoning and with medium reasoning; Google Gemini~3 Flash at low and high thinking levels; DeepSeek Chat and DeepSeek Reasoner; and Anthropic Claude Opus~4.6 and Sonnet~4.6, each without and with extended thinking.\footnote{The API model identifiers are \texttt{gpt-5.2}, \texttt{gemini-3-flash-preview}, \texttt{deepseek-chat}, \texttt{deepseek-reasoner}, \texttt{claude-opus-4-6}, and \texttt{claude-sonnet-4-6}. The test suites were generated through the respective provider APIs between March and June~2026.} Treating each reasoning level as a distinct configuration allows us to separate reasoning effects from prompt effects.

\subsection{LLMs Test-Generation Prompts}\label{sec:llm-prompts}

Prompts determine the context, constraints, and output format presented to an LLM. 
In test generation, they can specify the unit under test, testing framework, coverage goals, edge cases, dependency handling, and required response format.
Every strategy in our experiment receives the same extracted source-code context and targets the same \texttt{pytest} output. 
The strategies differ only in their instructions: the proposed prompt includes explicit analysis and mutation-resistance guidance, while the three literature baselines use shorter zero-shot or few-shot formulations. 
This controlled design allows observed differences to be attributed to prompt wording rather than unequal information.

\subsection{Test Evaluation Metrics}\label{sec:eval-metrics}

We evaluate generated suites along four orthogonal dimensions: fault coverage (fault-detection effectiveness), test execution (whether a suite runs at all), structural coverage (test adequacy), and test-quality indicators (maintainability). 
Each metric is defined once below and reused without restatement elsewhere in the paper.

\subsubsection{Fault Coverage: Mutation Testing}\label{sec:def-mutation}

Coverage metrics \cite{al2021jacoco} measure code execution but provide no guarantee that a test suite will detect actual faults \cite{inozemtseva2014coverage}. 
Mutation testing addresses this limitation by systematically assessing whether tests can distinguish correct programs from faulty variants.

For a Python implementation, Cosmic Ray derives mutants by applying small syntactic changes. 
Its default operators include arithmetic and relational operator replacement, conditional replacement, statement deletion, and constant replacement. 
We use the complete default operator set without modification and restrict the analysis to first-order mutants.
The mutation score of a test suite \textit{T} against a collection of mutants, as determined by Cosmic Ray, is the ratio of number of mutants killed by \textit{T}.

\noindent A mutant is killed when the test suite distinguishes it from the original implementation. Further details about the mutation engine and its operators appear in the Cosmic Ray documentation~\citep{cosmicray}.

\subsubsection{Test Execution: Executable Test Percentage}\label{sec:def-exec}

Generated tests must be executable before any coverage or fault-detection number is meaningful. We treat execution as a first-class outcome. Each generated test file must first parse without \texttt{SyntaxError} exceptions; this binary syntax-validity check is a prerequisite for all subsequent analyses. For each (model, prompt) cell we then record the \emph{Executable Test Percentage}: the proportion of ClassEval classes for which the generated suite ran Cosmic Ray mutation testing to completion (\texttt{cosmicray\_status == "success"}):

\begin{equation}\label{eq:exec-pct}
    \text{Executable Test Percentage} = \frac{|\text{Classes with Cosmic Ray success}|}{|\text{Classes attempted}|} \times 100\%
\end{equation}

\noindent A class that fails the gate typically reflects a suite that crashed during collection, raised an import error, or exceeded the mutation-testing timeout. No imputation is applied; failing cells are dropped from the executable subset rather than scored as zero.

\subsubsection{Structural Coverage}\label{sec:def-coverage}

Test metrics quantify how thoroughly a suite exercises the code under test~\citep{al2018using}. We use \texttt{coverage.py} to compute two measures. \emph{Line coverage} is the proportion of executable source lines run at least once:

\begin{equation}\label{eq:line-cov}
    \text{Line Coverage} = \frac{|\text{Executed Lines}|}{|\text{Executable Lines}|} \times 100\%
\end{equation}

\noindent Line coverage is a weak adequacy criterion in isolation: a suite may achieve high line coverage while exercising only a single path through conditional logic. \emph{Branch coverage} strengthens this criterion by quantifying the proportion of control-flow decision outcomes (both true and false branches) that are exercised:

\begin{equation}\label{eq:branch-cov}
    \text{Branch Coverage} = \frac{|\text{Executed Branches}|}{|\text{Total Branches}|} \times 100\%
\end{equation}

\subsubsection{Test-Quality Indicators}\label{sec:def-quality}

The practical utility of generated tests depends not only on fault detection but also on maintainability. We employ the \texttt{radon} static-analysis tool to compute the following indicators of test code.

\paragraph{Cyclomatic Complexity.} Cyclomatic complexity measures the number of linearly independent paths through test code:
\begin{equation}\label{eq:cyclomatic}
    V(G) = E - N + 2P
\end{equation}
\noindent where $E$, $N$, and $P$ are the numbers of edges, nodes, and connected components of the control-flow graph. High complexity in test code suggests overly intricate logic that may itself become a source of errors.

\noindent\textbf{Halstead metrics.} We compute program volume, difficulty, and effort from counts of unique operators and operands~\citep{halstead}. These measures provide a complementary view of the effort required to understand and maintain generated tests.

\noindent\textbf{Test smells.} Test smells are recurring anti-patterns, such as Assertion Roulette, magic numbers, and duplicate logic, that hinder comprehension and maintenance. They occur frequently in LLM-generated Python tests~\citep{alves2024detecting}. We do not quantify test smells; instead, we report assertion density, complexity, and suite size as measurable structural indicators.

\section{Related Work}\label{sec:relatedwork}

\begin{table*}[htbp]
\centering
\caption{Summary of related LLM-based test generation studies. Coverage abbreviations: L = Line, B = Branch, M = Method, P = Path. {Level = granularity of the unit under test (Func.\ = function/API, Meth.\ = method, Class, Proj.\ = project). Lang.\ = target language.} Mut.\ = Mutation testing used.}
\label{tab:llm_studies}
\footnotesize
\setlength{\tabcolsep}{4pt}
\begin{tabularx}{\textwidth}{@{}
  >{\raggedright\arraybackslash\hsize=0.85\hsize}X  
  >{\raggedright\arraybackslash\hsize=1.55\hsize}X  
  >{\raggedright\arraybackslash\hsize=1.05\hsize}X  
  r                                                 
  >{\raggedright\arraybackslash\hsize=0.70\hsize}X  
  >{\raggedright\arraybackslash\hsize=0.85\hsize}X  
  c                                                 
@{}}
\toprule
\textbf{Study} & \textbf{LLM(s)} & \textbf{Prompt Strategy} & \textbf{Sample} & {\textbf{Lang./Level}} & \textbf{Metrics} & \textbf{Mut.} \\
\midrule
Wang et al.~\cite{wang2025testeval} & GPT, Gemini, CodeLlama, Llama3, DeepSeek & Zero-shot, CoT & 210 & {Py / Func.} & L, B, P & No \\
Ouedraogo et al.~\cite{ouedraogo2024large} & GPT-3.5, GPT-4, Mistral, Mixtral & ZSL, FSL, CoT, ToT, GToT & 30 & {Py / Func.} & L, B, P & No \\
Chudic \& \c{C}al{\i}kl{\i}~\cite{chudic2026automated} & GPT-4o & Few-shot & 25 & {Py / Class} & L, B & No \\
{Sch\"afer et al.~\cite{schafer2024empirical} (TestPilot)} & {GPT-3.5, Codex, StarCoder} & {Zero-shot + docs} & {1{,}684 fns} & {JS / Func.} & {L, B} & {No} \\
Wang et al.~\cite{wang2025projecttest} & 9 frontier LLMs & Zero-shot & 60 projects & {Multi / Proj.} & L, B, M & No \\
Pan et al.~\cite{pan2025aster} & 6 models (GPT-4-turbo, Llama3, etc.) & Static analysis guided & 283 modules & {Java,Py / Meth.} & L, B, M & No \\
Chen et al.~\cite{chen2024chatunitest} & GPT-3.5-turbo & Adaptive focal context & 264 methods & {Java / Meth.} & L & No \\
Alves et al.~\cite{alves2024detecting} & GitHub Copilot & NL with scenario specs & 194 tests & {Py / Meth.} & Test smells & No \\
D\'{i}az-Arrieta et al.~\cite{diazarrieta2025comparative} & GPT-4o & Expert role prompt & 7 blocks & {Py / Mixed} & L, Faults & No \\
Yang et al.~\cite{yang2025rted} & DeepSeek-V3 & Multi-agent reflective & 69 errors & {Py / Meth.} & Type errors & No \\
Guilherme et al.~\cite{guilherme2023initial} & GPT-3.5, GPT-4 & Zero-shot & 13 classes & {Java / Class} & L, B & Yes \\
Abdullin et al.~\cite{abdullin2025test} & LLM vs.\ SBST vs.\ symbolic & Multiple & varies & {Java / Meth.} & L, B & Yes \\
{Wang et al.~\cite{wang2025mutgen} (MutGen)} & {LLM (GPT-4o)} & {Mutation-guided} & {204 subj.} & {Java / Meth.} & {Mutation} & {Yes} \\
{Alshahwan et al.~\cite{metaach2026} (Meta ACH)} & {LLM} & {Mutation-guided} & {10{,}795 cls} & {Kotlin / Class} & {Mutation} & {Yes} \\
\midrule
\textbf{This study} & \textbf{GPT-5.2, Gemini~3, DeepSeek{, Claude}$^{\dagger}$} & \textbf{4 prompts compared} & \textbf{100 classes} & {\textbf{Py / Class}} & \textbf{L, B, Mutation, Complexity} & \textbf{Yes} \\
\bottomrule
\end{tabularx}
{\footnotesize $^{\dagger}$ Ten configurations across four providers are reported: six from OpenAI, Google, and DeepSeek, and four Anthropic Claude configurations---Opus~4.6 and Sonnet~4.6, each without and with extended thinking (see \autoref{sec:benchmark-setup}).}
\end{table*}

Research on LLM-based test generation has grown rapidly, yet the literature exhibits systematic gaps in evaluation methodology, prompt comparison, and benchmark scale. We organize prior work along four threads: the breadth of LLM-based test-generation studies, prompting strategies, how evaluation approaches compare (including the rare cases that use mutation testing and the granularity at which units are tested), and the resulting research gap.

\subsection{LLM-Based Test Generation Studies}\label{sec:rw-studies}

LLM-based unit-test generation has been studied across languages, models, and tool forms. \citet{wang2025projecttest} evaluate nine frontier LLMs on 60 projects using compilation rate, correctness rate, and line coverage. \citet{pan2025aster} assess ASTER across Java and Python with line, branch, and method coverage; the system performs competitively with EvoSuite on Java SE and improves coverage on Java EE. \citet{chen2024chatunitest} evaluate ChatUniTest on 264 Java methods and report 59.6\% line coverage, outperforming TestSpark and EvoSuite. TestEval provides standardized targets for line, branch, and path coverage~\citep{wang2025testeval}.

TestPilot generates tests for JavaScript API functions from documentation-augmented prompts and reports median statement coverage of 70.2\%~\citep{schafer2024empirical}. ChatTester was among the first systems to prompt an LLM for class-level tests and then repair them using compiler feedback~\citep{chattester}. Other studies examine LLM-generated JUnit tests~\citep{siddiq2024using}, a LLaMA-based JUnit generator~\citep{gorla2025cubetesterai}, and broader evaluation practices~\citep{yang2024evaluation}. Tool-oriented work includes Copilot studies~\citep{humalajoki2024unit,bjork2024comprehensibility}, end-to-end generation and assessment~\citep{lops2025system}, and automated test refactoring~\citep{gao2024automated}. Nafis et al. incorporate mutation testing into Python test generation, but evaluate only one prompt configuration~\citep{nafis2025llm}. Across this literature, fault detection is either omitted or measured for a single model--prompt setting. Related work has also proposed structural and textual coupling measures for class-level integration testing~\citep{alazzam2024integration}.

\subsection{Prompting Strategies for Test Generation}\label{sec:rw-prompting}

Prompt design varies widely, yet controlled comparisons remain scarce. \citet{ouedraogo2024large} evaluate five prompting techniques across four LLMs and 30 Python programs. Chain-of-thought prompting generally improves coverage, although the effect depends on the model. Their evaluation is limited to structural coverage.

Most other studies use one prompt strategy without a controlled comparison. \citet{wang2025testeval} study zero-shot and chain-of-thought prompts but do not isolate prompt effects from model effects. \citet{chudic2026automated} use few-shot prompting with GPT-4o, so model capability and prompt design remain confounded. \citet{yang2025rted} develop a reflective multi-agent framework for type-error detection; its specialized agents make gains difficult to attribute to prompt wording alone. Static-analysis context~\citep{pan2025aster}, coverage-guided prompts~\citep{ryan2024code,yang2024advancing}, and method slicing~\citep{wang2024hits} can improve coverage, but these approaches do not target fault detection.

\subsection{Comparison of Evaluation Approaches}\label{sec:rw-eval}

The literature shares a methodological tendency that motivates our framework: structural coverage dominates, while fault detection and executability are rarely measured. A smaller group of studies uses richer criteria. \citet{alves2024detecting} detect test smells in Copilot-generated tests and find anti-patterns in 47.4\% of 194 tests. Assertion Roulette is the most prevalent smell, appearing in 42\%. \citet{diazarrieta2025comparative} compare human and GPT-4o tests directly; manual tests detect significantly more faults despite similar coverage above 80\%. These results reinforce the distinction between coverage and fault detection.

Mutation testing remains rare, and existing studies use it in two different ways. \citet{guilherme2023initial} evaluate GPT-generated tests with mutation analysis, but cover only two models and 13 Java classes. \citet{abdullin2025test} compare LLM-based generation with search-based and symbolic methods without varying prompts. A Python study also uses mutation testing with a single prompt~\citep{nafis2025llm}. Mutation analysis has additionally been adapted to learned components, including autoencoders~\citep{alsobeh2025muae}.

A separate line of work uses mutation to guide generation rather than evaluate it. MutGen feeds surviving mutants back into the prompt~\citep{wang2025mutgen}; MuTAP injects mutation feedback into prompt-based learning~\citep{mutap}; and Meta's Automated Compliance Hardening generates area-specific mutants and matching tests for 10{,}795 Kotlin classes~\citep{metaach2026}. These systems use mutants as feedback within one generation pipeline. Our study instead fixes the generation procedure and uses Cosmic Ray as the comparative yardstick across a four-prompt $\times$ ten-model factorial. We leave direct comparison with mutation-guided generators to future work (\autoref{sec:future-work}).

Benchmark choice and granularity also constrain generalizability. TestEval and TestPilot target individual functions~\citep{wang2025testeval,schafer2024empirical}; function-level benchmarks omit state management, method interactions, and constructor dependencies. ClassEval provides 100 curated Python classes with human-written test suites~\citep{du2024classeval}. Other controlled studies use smaller samples, including seven code blocks~\citep{diazarrieta2025comparative}, 25 classes~\citep{chudic2026automated}, and 30 programs~\citep{ouedraogo2024large}. ProjectTest covers entire codebases but offers less control for isolating prompt effects~\citep{wang2025projecttest}. Language also matters: ChatUniTest, HITS, and MutGen target Java methods~\citep{chen2024chatunitest,wang2024hits,wang2025mutgen}, while TackleTest and Meta ACH reach class level in Java and Kotlin~\citep{tackletest,metaach2026}. Controlled mutation-based evaluation of class-level Python test generation, where constructors, state, and method dependencies interact, remains limited.

\subsection{Research Gap Analysis}\label{sec:rw-gap}

As \autoref{tab:llm_studies} shows, three gaps remain. First, structural coverage dominates evaluation; few studies use mutation testing, and none of those varies prompts across several models. Second, controlled prompt comparisons rely primarily on coverage metrics, which cannot distinguish the strategies in our data. Third, prior work does not combine a factorial prompt-by-model design with mutation-based evaluation of class-level Python code. We address these gaps by evaluating 100 classes under 40 conditions ($4$ prompts $\times$ $10$ model configurations), producing 4,000 generated suites for factorial analysis.

To our knowledge, this is the first study to compare several prompts and LLMs systematically through mutation testing at class level in Python. The language determines the mutation engine. PIT/PITest mutates JVM bytecode and is restricted to Java; Cosmic Ray performs source-level mutation on Python abstract syntax trees and can target an individual class. This design distinguishes our study from work on Java methods with MutGen~\citep{wang2025mutgen} and Kotlin classes with Meta ACH~\citep{metaach2026}, neither of which varies prompt and model in a controlled factorial.

\section{Research Methodology}\label{sec:methodology}

This section describes test generation and evaluation preparation. It begins with the assessment framework (\autoref{sec:assessment-framework}), then covers source-code context extraction, prompt design, benchmark prompts, the LLM benchmarking setup, and output post-processing. \autoref{sec:pipeline} integrates these components into the end-to-end workflow. Experimental outcomes appear in \autoref{sec:results}.

\subsection{Test-Suite Assessment Framework}\label{sec:assessment-framework}

We assess each generated suite along the four dimensions defined in \autoref{sec:eval-metrics}: fault coverage, execution (\autoref{eq:exec-pct}), structural coverage (Eqs.~\ref{eq:line-cov}--\ref{eq:branch-cov}), and test-quality indicators. No single measure is sufficient. Coverage records exercised paths; mutation score measures fault sensitivity, but only for suites that run; quality indicators address maintainability rather than effectiveness. A fixed toolchain applies \texttt{pytest}, \texttt{coverage.py}, Cosmic Ray, and \texttt{radon} uniformly to every model--prompt cell and to the human baseline. \autoref{sec:results} reports the numerical outcomes.

\subsection{Prompt Engineering and LLM-Based Test Case Generation}\label{sec:prompt-engineering}

The four prompt strategies evaluated in this study span a range of design philosophies and shot types (zero-shot, few-shot), enabling a controlled comparison of how prompt complexity affects test quality. The subsections below describe how source code is prepared for the model (\autoref{sec:context-extraction}), the proposed prompt and its rationale (\autoref{sec:promptdesign}), the benchmark prompt variants used for comparison (\autoref{sec:benchmark-prompts}), the LLM-and-prompt benchmarking setup (\autoref{sec:benchmark-setup}), and how raw model output is turned into executable test files (\autoref{sec:postproc}).

\subsubsection{Source Code Context Extraction}\label{sec:context-extraction}

The first stage prepares source code for insertion into every prompt. It parses ClassEval files to extract complete class definitions, fully qualified module names, parameter signatures, type annotations, imported dependencies, and known edge cases. This context helps the model generate valid imports, mock external dependencies, and cover boundaries. Every prompt receives the same extracted information, so differences in output reflect instructions rather than unequal context.

\subsubsection{Proposed Prompt Design}\label{sec:promptdesign}

The proposed prompt is designed around a single hypothesis: that guiding the model through explicit analysis and planning \emph{before} code emission, while stating mutation-resistance and quality objectives directly, yields more fault-sensitive suites than terse instructions. Concretely, the prompt is composed of seven components.

\begin{itemize}
    \item \textbf{Role.} A system role casts the model as an expert Python test engineer specializing in comprehensive coverage, anchoring tone and rigor.
    \item \textbf{Inserted source-code context.} The extracted class definition, module path, dependencies, and known edge cases (\autoref{sec:context-extraction}) are injected via template variables so the model tests the actual unit, not a paraphrase.
    \item \textbf{Testing framework and output format.} The prompt fixes \texttt{pytest} as the framework and requires a single, complete, directly executable test file with all imports and no surrounding prose or markdown.
    \item \textbf{Executable-test requirements.} Structural requirements (fixtures for setup, parametrization for similar cases, descriptive names, independent tests, specific assertions) push the output toward suites that import and collect cleanly.
    \item \textbf{Explicit constraints.} Quality standards forbid magic numbers, \texttt{time.sleep()} calls, trivial assertions, and duplicate logic, and require one behavior per test.
    \item \textbf{Normal, edge, and failure coverage.} A six-step analysis phase enumerates execution paths, public/private methods, constructor ranges, dependencies to mock, exception scenarios, and state invariants, after which the model is asked to cover normal cases, edge cases, and error cases for each method.
    \item \textbf{Mutation-resistance instructions.} The prompt explicitly asks for tests that fail if an operator, constant, conditional, or return value changes, thereby directing the model toward discriminating assertions rather than mere path execution.
\end{itemize}

The last two components are what most distinguish the proposed prompt from the baselines of \autoref{sec:benchmark-prompts}: none of the three literature prompts contains an explicit analysis phase or mutation-resistance instructions (see \autoref{tab:prompt_comparison}). The full prompt is shown below.

\begin{tcolorbox}[
    enhanced,
    breakable,
    fonttitle=\bfseries\tiny,
    fontupper=\tiny,
    arc=3mm,
    boxrule=1.0pt,
    left=6pt, right=6pt, top=6pt, bottom=6pt
]
\textbf{\textcolor{blue}{SYSTEM ROLE:}}

You are an expert Python test engineer specializing in comprehensive test coverage.

\vspace{0.3cm}
\textbf{\textcolor{blue}{TASK:}}

Generate a complete pytest test suite for the following Python \{class/function\}.

\vspace{0.3cm}
\textbf{\textcolor{blue}{COVERAGE OBJECTIVES (Priority Order):}}
\begin{enumerate}
    \item 100\% line coverage - execute every statement
    \item 100\% branch coverage - test both True/False for all conditionals
    \item 100\% method coverage - call every method at least once
    \item Mutation resistance - tests should detect code changes
    \item Edge case coverage - boundaries, None, empty, negative values
\end{enumerate}

\vspace{0.3cm}
\textbf{\textcolor{blue}{ANALYSIS PHASE:}}

Before generating tests, analyze:
\begin{itemize}
    \item All execution paths and branches
    \item All methods (public and private)
    \item Constructor parameters and valid ranges
    \item Dependencies requiring mocking
    \item Exception scenarios (raise statements, try/except)
    \item State transitions and invariants
\end{itemize}

\vspace{0.3cm}
\textbf{\textcolor{blue}{TEST GENERATION REQUIREMENTS:}}
\begin{enumerate}
    \item \textbf{Framework:} Use pytest
    \item \textbf{Imports:} from \{module\_path\} import \{class\_name\}
    \item \textbf{Structure:}
    \begin{itemize}
        \item One test file: test\_\{class\_name\}.py
        \item Use fixtures for complex setup (@pytest.fixture)
        \item Use parametrize for similar test cases
        \item Group related tests in classes (optional)
    \end{itemize}
    \item \textbf{Test Case Design:}
    \begin{itemize}
        \item Descriptive names: test\_<method>\_<scenario>\_<expected>
        \item Independent tests (no shared state)
        \item Specific assertions with messages
        \item Cover: normal cases, edge cases, error cases
    \end{itemize}
    \item \textbf{Quality Standards:}
    \begin{itemize}
        \item No magic numbers (use variables)
        \item No time.sleep() calls
        \item Clear assertion messages: assert x == 5, "Expected 5 items"
        \item Mock external dependencies
        \item Each test validates ONE specific behavior
    \end{itemize}
    \item \textbf{Mutation Detection:}

    Generate tests that will FAIL if:
    \begin{itemize}
        \item Any operator is changed (+/-, ==/<, and/or)
        \item Any constant is modified
        \item Any conditional is inverted
        \item Return values change
    \end{itemize}
\end{enumerate}

\vspace{0.3cm}
\textbf{\textcolor{blue}{AVOID These Test Smells:}}
\begin{itemize}
    \item Multiple assertions without messages
    \item Unexplained numeric literals
    \item Trivial assertions (assert True)
    \item Duplicate test logic
\end{itemize}

\vspace{0.3cm}
\textbf{\textcolor{blue}{OUTPUT FORMAT:}}
\begin{itemize}
    \item Only Python code (no markdown, no explanations)
    \item Complete, executable test file
    \item All necessary imports included
    \item Proper pytest structure
\end{itemize}

\vspace{0.3cm}
\textbf{\textcolor{blue}{CODE UNDER TEST:}}

\{source\_code\}

\vspace{0.3cm}
\textbf{\textcolor{blue}{ADDITIONAL CONTEXT:}}
\begin{itemize}
    \item Module path: \{module\_path\}
    \item Dependencies: \{dependencies\}
    \item Known edge cases: \{edge\_cases\}
\end{itemize}
\vspace{0.3cm}
\end{tcolorbox}

\noindent\textbf{Reasoning strategy: zero-shot with embedded chain-of-thought.} Previous studies use either zero-shot prompting~\citep{alves2024detecting,wang2025projecttest,diazarrieta2025comparative,pan2025aster,chen2024chatunitest} or few-shot examples~\citep{yang2025rted,wang2025testeval}. Our prompt uses a three-phase structure (analysis $\rightarrow$ planning $\rightarrow$ generation) within a single zero-shot instruction. It does not place example tests in the context window. This design limits context overhead while eliciting structured reasoning. The scaffolding appears below.

\begin{tcolorbox}[
    enhanced, breakable, fonttitle=\bfseries,
    title={Reasoning Strategy: Zero-Shot with Embedded Chain-of-Thought},
    arc=1mm, boxrule=1pt, left=4pt, right=4pt, top=4pt, bottom=4pt, fontupper=\tiny
]
\textbf{\textcolor{blue}{PHASE 1: ANALYSIS}}

Think step-by-step and identify:
\begin{enumerate}
    \item Count total methods: \underline{\hspace{3cm}}
    \item Count total branches: \underline{\hspace{3cm}}
    \item List edge cases: \underline{\hspace{3cm}}
    \item Identify dependencies: \underline{\hspace{3cm}}
\end{enumerate}

\vspace{0.3cm}
\textbf{\textcolor{blue}{PHASE 2: PLANNING}}

For each method, plan:
\begin{itemize}
    \item Normal case tests: \underline{\hspace{3cm}}
    \item Branch coverage tests: \underline{\hspace{3cm}}
    \item Edge case tests: \underline{\hspace{3cm}}
    \item Exception tests: \underline{\hspace{3cm}}
\end{itemize}

\vspace{0.3cm}
\textbf{\textcolor{blue}{PHASE 3: GENERATION}}

Generate the complete test suite following your plan.
Ensure each planned test case is implemented.
\end{tcolorbox}

\subsubsection{Prompt Templates and Benchmark Prompt Variants}\label{sec:benchmark-prompts}

To benchmark the proposed prompt against existing strategies, we adapted three prompt templates from the recent literature to work within our experimental framework. These are used strictly as \emph{baselines for comparison}, not as the proposed method. Each was modified to accept the same template variables (\texttt{\{source\_code\}}, \texttt{\{class\_name\}}, \texttt{\{module\_path\}}) while preserving its original structure and intent. \autoref{tab:prompt_comparison} summarizes the key design differences across all four prompts.

\begin{table*}[htbp]
\centering
\caption{Comparison of the four prompt strategies evaluated in this study, showing key design differences.}
\label{tab:prompt_comparison}
\footnotesize
\begin{tabular}{lcccc}
\toprule
\textbf{Feature} & \textbf{Ours} & \textbf{Wang et al.} & \textbf{Chudic \& \c{C}al{\i}kl{\i}} & \textbf{Ouedraogo et al.} \\
\midrule
Lines & 77 & 30 & 29 & 18 \\
Shot Type & Zero-shot CoT & Zero-shot & Few-shot style & Zero-shot \\
System Role & Expert test engineer & Task-focused & Expert with examples & Professional tester \\
Analysis Phase & Explicit 6-step & None & None & None \\
Coverage Target & 5-tier priority & Line + branch & High coverage & All methods \\
Mutation Focus & Explicit instructions & None & None & None \\
Output Format & Detailed requirements & Standard & Test class format & Marker boundaries \\
Citation & (this work) & \cite{wang2025testeval} & \cite{chudic2026automated} & \cite{ouedraogo2024large} \\
\bottomrule
\end{tabular}
\end{table*}

\noindent\textbf{Wang et al.\ (TestEval).} This 30-line prompt uses a structured task description with explicit line- and branch-coverage targets~\citep{wang2025testeval}. It provides delimited source code, specifies the import format, and asks for additional tests that exercise different paths. It does not include the analysis phases used in our prompt.

\begin{tcolorbox}[enhanced, breakable, fonttitle=\bfseries\small, fontupper=\small\ttfamily, arc=2mm, boxrule=0.5pt, left=4pt, right=4pt, top=4pt, bottom=4pt, title={Wang et al.\ (TestEval) Prompt Template}]
\textbf{TASK:}\\
Please write test methods for the \{class\_or\_function\} '\{class\_name\}' given the following program under test.\\[4pt]
\textbf{PROGRAM UNDER TEST:}\\
\{source\_code\}\\[4pt]
\textbf{KNOWN EDGE CASES:}\\
\{edge\_cases\}\\[4pt]
\textbf{COVERAGE TARGET:}\\
Your test cases must achieve full line and branch coverage of the \{class\_or\_function\} under test.\\[4pt]
\textbf{OUTPUT FORMAT:}\\
Your test file should use pytest and begin with:\\
from \{module\_path\} import \{class\_name\}\\[4pt]
- Only output Python code (no markdown, no explanations)\\
- Complete, executable test file with all necessary imports\\
- Each test method should validate one specific behavior\\[4pt]
\textbf{FOLLOW-UP:}\\
Generate additional test methods that cover different statements, branches, and edge cases. Your tests must be different from each other and should cover different execution paths.
\end{tcolorbox}

\noindent\textbf{Chudic and \c{C}al{\i}kl{\i}.} This 29-line prompt uses a few-shot framing that refers to example test cases~\citep{chudic2026automated}. The system message assigns expertise in Python test generation with \texttt{pytest} and asks for maintainable tests with high coverage. It provides structural guidance without an explicit analysis phase.

\begin{tcolorbox}[enhanced, breakable, fonttitle=\bfseries\small, fontupper=\small\ttfamily, arc=2mm, boxrule=0.5pt, left=4pt, right=4pt, top=4pt, bottom=4pt, title={Chudic \& \c{C}al{\i}kl{\i} Prompt Template}]
\textbf{SYSTEM:}\\
You are an expert in Python test generation using pytest. Your goal is to generate new high-quality unit tests for a given Python class. You will be provided with the class definition and your output should be a list of new unit tests. The prompt will include EXAMPLES of similar test cases to help you generate well-structured test cases. Make sure to keep the tests maintainable and easy to understand, while aiming for high code coverage. The output should only include the test classes.\\[4pt]
\textbf{USER:}\\
\# CLASS UNDER TEST: \{class\_name\}\\
\{source\_code\}\\[4pt]
\textbf{KNOWN EDGE CASES:}\\
\{edge\_cases\}\\[4pt]
\textbf{DEPENDENCIES:}\\
\{dependencies\}\\[4pt]
\textbf{OUTPUT FORMAT:}\\
Your test file should use pytest and begin with:\\
from \{module\_path\} import \{class\_name\}\\[4pt]
- Only output Python code (no markdown, no explanations)\\
- Complete, executable test file with all necessary imports\\
- Each test method should validate one specific behavior\\
- Aim for high line and branch coverage
\end{tcolorbox}

\noindent\textbf{Ouedraogo et al.} This 18-line prompt is the shortest strategy~\citep{ouedraogo2024large}. It assigns the role of professional software tester, requests comprehensive \texttt{pytest} cases for every method, and delimits the response with start and end markers. Its concise instructions leave the model greater discretion.

\begin{tcolorbox}[enhanced, breakable, fonttitle=\bfseries\small, fontupper=\small\ttfamily, arc=2mm, boxrule=0.5pt, left=4pt, right=4pt, top=4pt, bottom=4pt, title={Ouedraogo et al.\ Prompt Template}]
\textbf{USER:}\\
As a professional software tester who writes Python test methods, generate pytest test cases to comprehensively test all methods in the following class named \{class\_name\}. The complete pytest test file must start with \#\#\#Test START\#\# and finish with \#\#\#Test END\#\#\\
Here is the \{class\_name\} class:\\
\{source\_code\}\\[4pt]
\textbf{OUTPUT FORMAT:}\\
Your test file should use pytest and begin with:\\
from \{module\_path\} import \{class\_name\}\\[4pt]
- Only output Python code (no markdown, no explanations)\\
- Complete, executable test file with all necessary imports\\
- Each test method should validate one specific behavior
\end{tcolorbox}

\subsubsection{LLMs and Prompt Benchmarking Setup}\label{sec:benchmark-setup}

The benchmark crosses the proposed prompt with three baseline prompts and applies all four to the LLM configurations defined in \autoref{sec:llm-thinking}. The factorial contains ten configurations from four providers: OpenAI GPT-5.2 (no reasoning and medium reasoning), Google Gemini~3 Flash (low and high thinking), DeepSeek Chat and Reasoner, and Claude Opus~4.6 and Sonnet~4.6 (each without and with extended thinking). This design yields $4 \times 10 = 40$ cells, each evaluated on all 100 ClassEval classes.
Comparing multiple model--prompt configurations is essential to the benchmark's purpose: it lets us attribute variation in test quality to the prompt, to the model, or to their reasoning levels, rather than confounding them as prior single-model or single-prompt studies do (\autoref{sec:relatedwork}).

\subsubsection{{Scope: Considered Test-Generation Tools and Models}}\label{sec:tool-scope}

Many LLM-based test-generation tools exist, but including them all would confound the factors this study isolates. \autoref{tab:tool-scope} records the systems considered and the decision for each. Three principles guide the scope. First, our subjects are class-level Python, so tools for Java, JavaScript, or Kotlin are discussed rather than executed. Porting them would change the pipeline, not only the prompt. Second, systems that bundle repair loops, slicing, or static analysis entangle prompt effects with pipeline engineering; we evaluate prompting of base LLMs directly. Third, closed chat interfaces lack documented batch APIs for controlled prompt injection and fixed sampling over 400 cases per model. Mutation-guided generators pursue a different objective because mutants drive generation rather than serve only as the evaluation measure. We leave direct comparisons with those systems to future work (\autoref{sec:future-work}).

\begin{table*}[htbp]
\centering
\caption{{Test-generation tools and systems considered, with the rationale for inclusion or exclusion. Level: granularity of the unit under test. ``Mut.'': uses mutation testing (guided = mutants drive generation; eval = mutants measure quality).}}
\label{tab:tool-scope}
\footnotesize
{%
\begin{tabular}{p{2.7cm} l l l l p{6.0cm}}
\toprule
\textbf{Tool / system} & \textbf{LLM} & \textbf{Lang.} & \textbf{Level} & \textbf{Mut.} & \textbf{Decision and rationale} \\
\midrule
TestPilot~\citep{schafer2024empirical} & Yes & JS & Func. & No & Excluded; discussed. JavaScript, function-level, and coverage-only; outside our class-level Python scope. \\
ChatUniTest / ChatTester~\citep{chen2024chatunitest,chattester} & Yes & Java & Meth./Class & No & Excluded; discussed. Java; bundles a generation--validation--repair loop that confounds the prompt effect we isolate. \\
HITS~\citep{wang2024hits} & Yes & Java & Meth. & No & Excluded; discussed. Java method-slicing technique; method-level; slicing confounds prompt attribution. \\
TackleTest\newline\citep{tackletest} & No & Java & Class & No (comb.) & Excluded; discussed. Search/combinatorial generator built on EvoSuite/Randoop; not an LLM prompt tool; Java. \\
JAIPilot & Yes & Java & Class/Meth. & No & Excluded; discussed. IntelliJ plugin (Gemini agent) with an interactive, self-correcting loop; not a reproducible Python batch harness. \\
MutGen~\citep{wang2025mutgen} & Yes & Java & Meth. & guided & Excluded; future head-to-head. Mutation-\emph{guided} generation (different objective); Java. \\
MuTAP~\citep{mutap} & Yes & Python & Func. & guided & Excluded; future head-to-head. Mutation-guided prompt learning at function level. \\
Meta ACH~\citep{metaach2026} & Yes & Kotlin & Class & guided & Excluded; discussed. Proprietary industrial pipeline; Kotlin; mutation-guided for compliance. \\
YesChat AI, Blackbox AI, Continuous Dev & Yes & n/a & n/a & No & Excluded. Closed web chat UIs with no documented batch API for controlled, reproducible prompt injection and fixed sampling. \\
This work (prompted base LLMs) & Yes & Python & Class & eval & Proposed prompt + three baseline prompts $\times$ ten model configurations; mutation score as the evaluation yardstick. \\
\bottomrule
\end{tabular}}

\footnotesize\textit{Tools references:\\} JAIPilot, \url{https://www.jaipilot.com/}; YesChat AI, \url{https://www.yeschat.ai/}; Blackbox AI, \url{https://www.blackbox.ai/}
\end{table*}

\subsubsection{Output Post-Processing}\label{sec:postproc}

LLM outputs often contain artifacts that prevent execution, including Markdown fences, explanatory prose, and inconsistent indentation. Post-processing extracts code, removes non-code text and fences, normalizes formatting, and validates the result with Python's abstract syntax tree parser. Code that fails to parse is recorded as a syntax failure. Valid code is written to a \texttt{test\_*.py} file, checked for importability against the target ClassEval module, and passed to the execution stage.

\subsubsection{Automated Test Case Generation Pipeline}\label{sec:pipeline}

\begin{figure}[htbp]
\centering
\includegraphics[width=\textwidth]{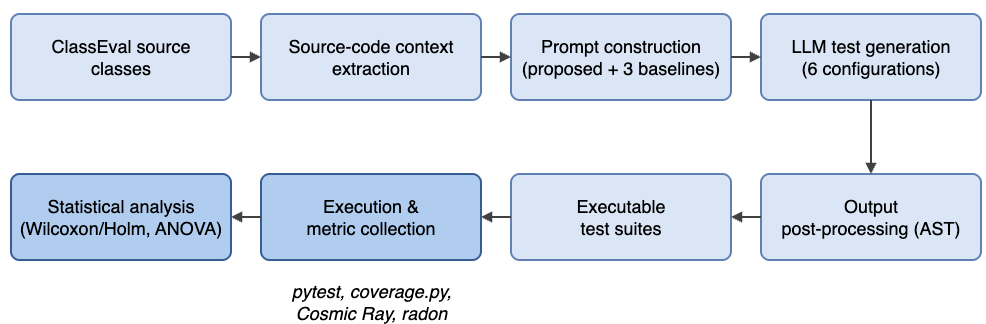}
\caption{{End-to-end automated pipeline for LLM-based test generation and evaluation. The generation stages (top row, light blue) produce executable suites; the evaluation stages (bottom-left, darker) consume them. }}
\label{fig:pipeline}
\end{figure}

An automated pipeline orchestrates the workflow across all experimental runs. As \autoref{fig:pipeline} shows, the stages are source-code context extraction, prompt construction, LLM-based test generation, output post-processing, and evaluation preparation. This subsection describes generation; \autoref{sec:results} reports the resulting evaluation.

Algorithm~\ref{alg:pipeline} states the workflow at a high level. Mutation score and the other metrics are defined once in \autoref{sec:eval-metrics} and are not re-derived here; the algorithm refers to them rather than recomputing their formulas.

\begin{algorithm}[htbp]
\caption{{End-to-End LLM-Based Test Case Generation and Evaluation Pipeline}}
\label{alg:pipeline}
\footnotesize
\textbf{Input:} source code $S$ and target units under test; prompt set $P=\{\text{Ours},\text{Wang},\text{Chudic},\text{Ouedraogo}\}$; LLMs configuration set $M$ (6 configs); human baseline $T_H$.\\
\textbf{Output:} validated test suites and evaluation results.
\begin{enumerate}[leftmargin=2.2em,itemsep=1pt,topsep=3pt]
  \item Extract source-code context $C \gets \textsc{ExtractContext}(S)$ (class definition, imports, signatures, dependencies, edge cases).
  \item \textbf{for each} prompt $p \in P$ and configuration $m \in M$ \textbf{do}
  \begin{enumerate}[label=2.\arabic*,leftmargin=2.4em,itemsep=0pt,topsep=1pt]
    \item Construct the prompt from $p$ and $C$ (proposed design or benchmark variant).
    \item Generate candidate test code with model $m$.
    \item Post-process: strip non-code text, AST-validate, write \texttt{test\_*.py}.
    \item \textbf{if} the suite executes (Cosmic Ray completes, Eq.~\ref{eq:exec-pct}) \textbf{then} record line/branch coverage (Eqs.~\ref{eq:line-cov}--\ref{eq:branch-cov}), mutation score (from \texttt{CosmicRay}), complexity (Eq.~\ref{eq:cyclomatic}), test smells, and execution status; \textbf{else} mark the cell non-executable and exclude it from the executable subset (no imputation).
  \end{enumerate}
  \item Compute $T_H$ metrics on its executable subset for the baseline comparison.
  \item Apply Wilcoxon signed-rank tests with Holm-Bonferroni correction across the reported metrics.
  \item Fit a two-way ANOVA (model, prompt) on the executable rows; run per-model Friedman omnibus tests.
  \item Rank and compare (model, prompt) configurations on the evaluation metrics.
  \item \textbf{return} validated suites and evaluation results.
\end{enumerate}
\end{algorithm}

\section{Experimental Evaluation}\label{sec:results}

This section reports the experimental evaluation. We first describe the experimental protocol and statistical analysis (\autoref{sec:experimental-setup}), then present results by dimension: suite executability and test execution (\autoref{sec:executability}), structural coverage (\autoref{sec:structural-coverage}), mutation testing (\autoref{sec:mutation-results}), test quality (\autoref{sec:test-quality}), the correlation between metrics (\autoref{sec:correlation}), and a comparative synthesis across LLMs and prompts (\autoref{sec:cross-prompt-synthesis}). Suite executability is treated as a first-class outcome because it conditions every downstream comparison: a test suite that never runs yields no coverage number and no mutation score.

\subsection{Experimental Protocol and Statistical Analysis}\label{sec:experimental-setup}

We use ClassEval, a benchmark of 100 curated Python classes with varied method counts, dependencies, and state-management patterns~\citep{du2024classeval}.\footnote{\url{https://github.com/FudanSELab/ClassEval}} We cross four prompts (ours and three literature baselines~\citep{wang2025testeval,chudic2026automated,ouedraogo2024large}) with ten LLM configurations from four providers. The 100 human-written ClassEval suites form the baseline. We execute suites with \texttt{pytest}, measure structural coverage with \texttt{coverage.py}, run mutation testing with Cosmic Ray, and calculate complexity with \texttt{radon}. The design produces $4 \times 10 \times 100 = 4{,}000$ LLM-generated suites plus 100 human-written suites.

The primary protocol treats execution as a gate. A class enters a model--prompt cell only when Cosmic Ray finishes (\texttt{cosmicray\_status == "success"}). Pairwise Wilcoxon tests use the intersection of executable classes, and group summaries use each cell's own executable subset. The executable-gated analyses do not impute failed suites. A separate Friedman sensitivity analysis aligns all 100 classes and assigns zero to non-executable suites; it therefore combines executability with mutation effectiveness and is interpreted separately.

\paragraph{{Statistical analysis.}}
Per-class comparisons are paired and non-normal, so we use Wilcoxon signed-rank tests for line coverage, branch coverage, and mutation score on shared executable classes. Pairs with fewer than ten shared classes are excluded. The dominance analysis applies Holm correction across the eligible mutation-score comparisons and reports adjusted values ($p_{\mathrm{adj}}$). We also fit a Type-II two-way analysis of variance (factors: model and prompt) to the executable mutation-score rows and report classical $\eta^2$. Finally, the zero-filled Friedman sensitivity analysis tests the joint effect of prompt choice and non-execution within each model; it is not an executable-subset estimate.

\begin{figure}[htbp]
\centering
\includegraphics[width=\textwidth]{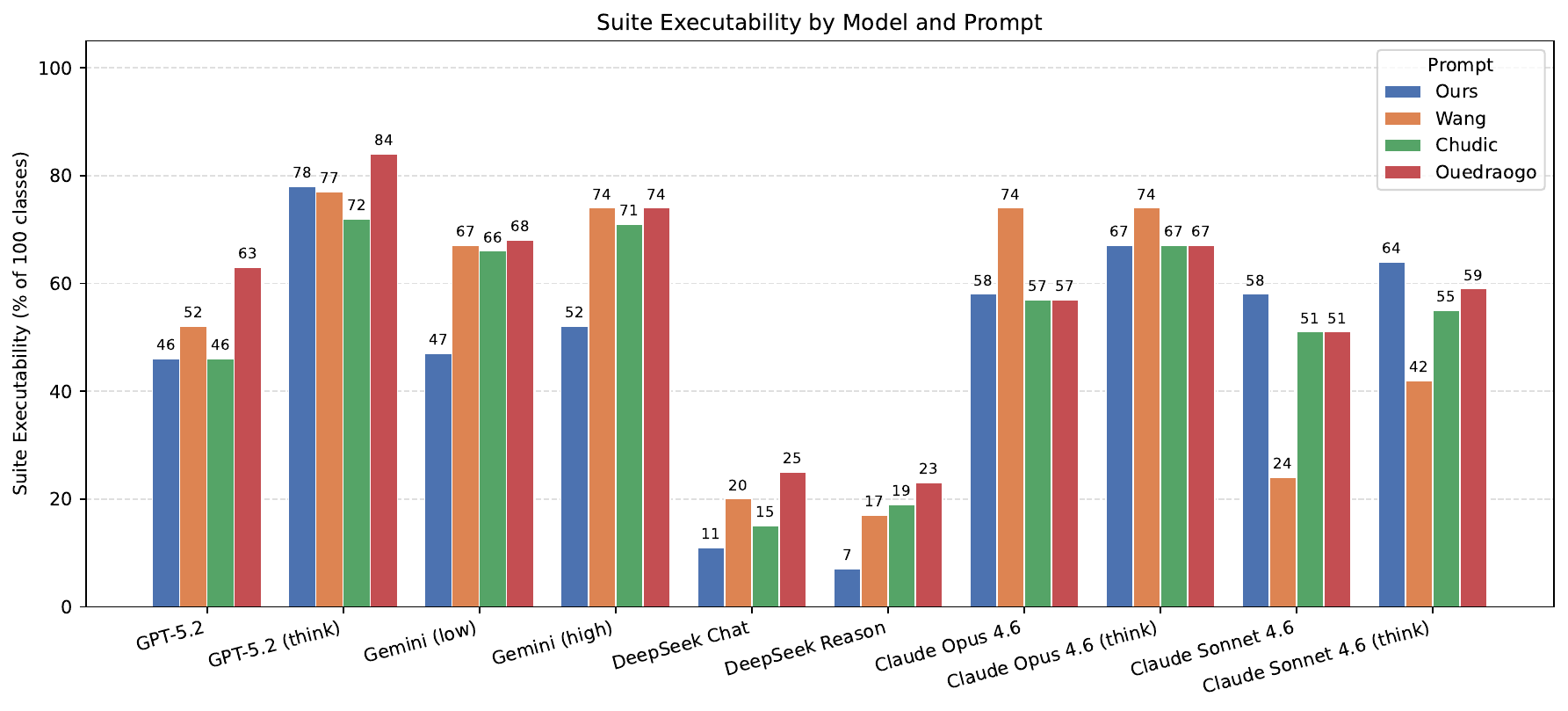}
\caption{Executable Test Percentage (\%) per (model, prompt) cell. Height is the count of ClassEval classes for which the generated suite completed Cosmic Ray mutation testing (out of 100). DeepSeek suites rarely execute; GPT-5.2 with medium reasoning executes most reliably.}
\label{fig:executable-proportion}
\end{figure}

\subsection{Suite Executability and Test Execution}\label{sec:executability}

xecutability varies by more than an order of magnitude across configurations, and the variation is dominated by model choice rather than prompt choice. \autoref{fig:executable-proportion} plots the proportion of the 100 ClassEval classes for which Cosmic Ray ran to completion, grouped by model and prompt. GPT-5.2 with medium reasoning is the most reliable generator: it clears the filter on 72 to 84 classes depending on prompt, with an overall rate of 77.8\%. DeepSeek is the least reliable. DeepSeek Chat executes on only 11 to 25 of 100 classes across the four prompts (overall 17.8\%), and DeepSeek Reasoner fares similarly (7 to 23 classes, overall 16.5\%). The Gemini and GPT-5.2 (no reasoning) configurations sit between these poles at 52 to 78\% overall. 
Claude occupies the middle of the executability range. Opus reaches 61.5\% without thinking and 68.8\% with thinking; Sonnet reaches 46.0\% and 55.0\%, respectively. Most Claude failures share one cause: the model follows the literal \texttt{from \{module\} import \{class\_name\}} instruction and does not infer a corrected class name when the injected hint is mis-cased. The resulting import failures resemble dependency hallucinations reported in broader LLM-development workflows~\citep{alsobeh2025shadowplay}.
Across prompts, Ouedraogo et al.'s minimal template produces the highest executability rate ({57.1\% of the 1{,}000 model-class cells}), followed by {Wang (52.1\%), Chudic (51.9\%), and our proposed prompt (48.8\%)}.

Two implications follow. First, any comparison that averages over all 100 classes conflates generation quality with executability. A failed suite contributes a zero mutation score not because its assertions are weak but because they never ran. Second, the DeepSeek results that follow must be read in the light of small executable denominators. Where we report a DeepSeek mutation score, it is computed on the 7 to 25 classes the suite actually survived.

\begin{figure}[htbp]
\centering
{\includegraphics[width=\textwidth]{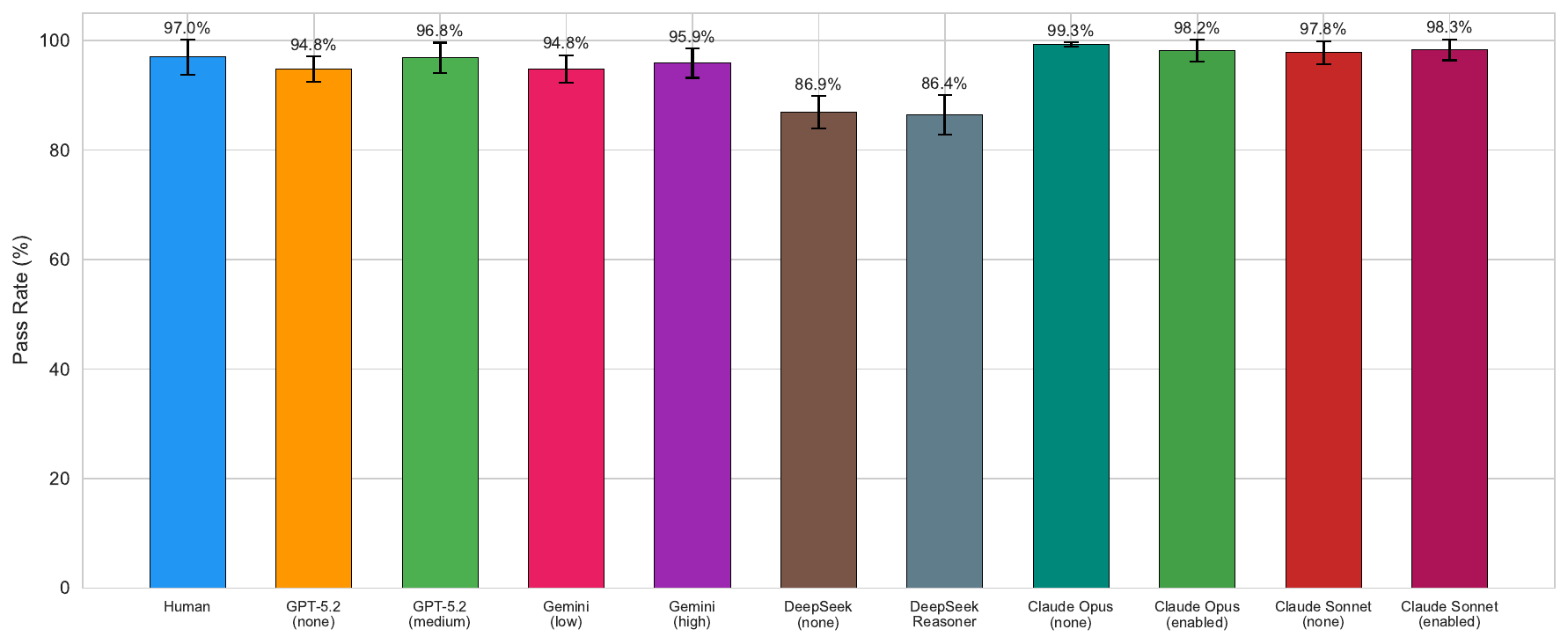}}
\caption{{Suite execution (pass) rate by model and prompt. The ordering mirrors \autoref{fig:executable-proportion}: GPT-5.2 (medium reasoning) and Gemini execute most reliably; DeepSeek configurations execute least often.}}
\label{fig:pass-rate}
\end{figure}

The same ordering appears when execution is summarized as a pass rate over generated suites: \autoref{fig:pass-rate} shows the proportion of suites that execute successfully by model and prompt, with DeepSeek configurations far below the others.

\subsection{Structural Coverage Results}\label{sec:structural-coverage}

\begin{figure}[htbp]
\centering
{\includegraphics[width=\textwidth]{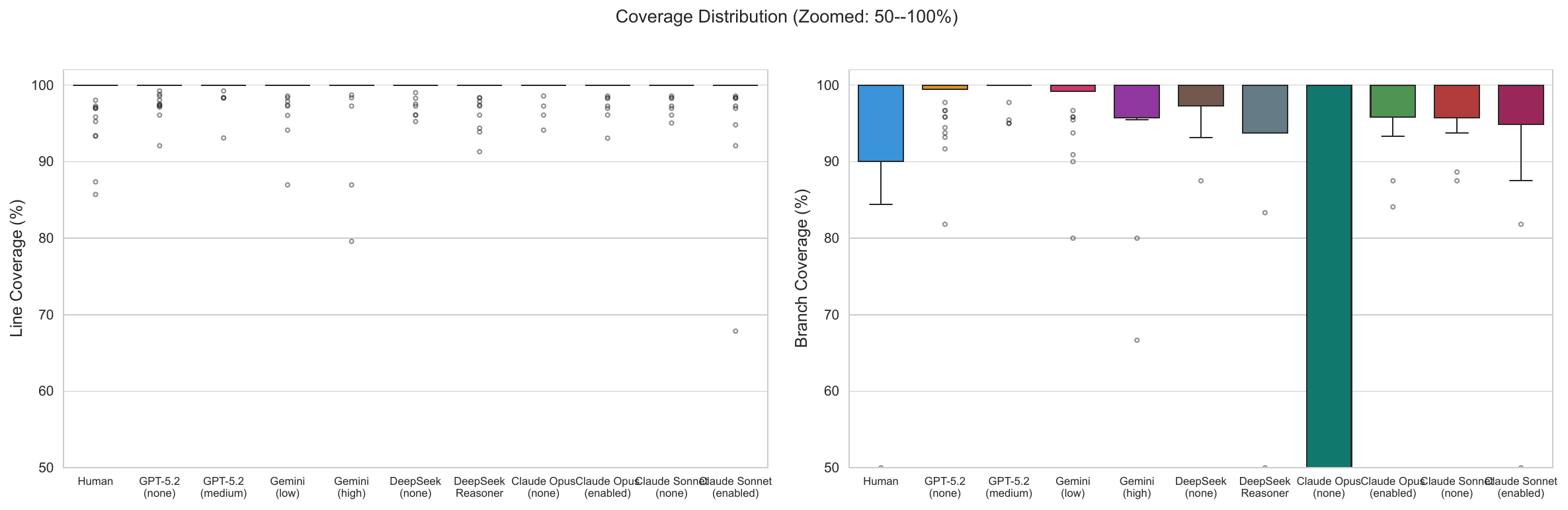}}
\caption{{Line and branch coverage distributions across (model, prompt) cells on executable classes (zoomed). Medians sit at 100\%; the near-ceiling spread confirms coverage cannot discriminate configurations.}}
\label{fig:coverage-box}
\end{figure}

Line coverage is at or near the ceiling in every (model, prompt) cell in the filtered data. Across the 40 cells, median line coverage is 100\% in every cell, while mean line coverage ranges from 99.32\% to 99.95\% in the 24 non-Claude cells and from 89.29\% to 98.49\% in the 16 Claude cells; the human baseline on its 82 executable rows is 99.37\% mean. Branch coverage follows the same pattern, ranging between 79.7\% and 95.3\% across cells with medians at 100\%. Coverage is therefore uninformative for ranking configurations in this dataset. For completeness, the per-cell values are reported in \autoref{tab:app-coverage} (relocated here from the former appendix), and the distribution of line and branch coverage across cells is shown in \autoref{fig:coverage-box}.

\begin{table*}[htbp]
\centering
\caption{Per-cell structural coverage on executable classes only. \textit{n} is the number of classes for which Cosmic Ray ran to completion for that (model, prompt) cell. Line and branch values are means; medians are 100\% for every cell.}
\label{tab:app-coverage}
\footnotesize
\begin{tabular}{lrrr}
\toprule
\textbf{Group} & \textbf{n} & \textbf{Mean Line (\%)} & \textbf{Mean Branch (\%)} \\
\midrule
Human                        & 82 & 99.37 & 91.06 \\
\midrule
GPT-5.2 / Ours               & 46 & 99.80 & 90.80 \\
GPT-5.2 / Wang               & 52 & 99.90 & 90.14 \\
GPT-5.2 / Chudi\v{c}         & 46 & 99.79 & 95.25 \\
GPT-5.2 / Ou\'edraogo        & 63 & 99.77 & 90.15 \\
\midrule
GPT-5.2 (think) / Ours       & 78 & 99.95 & 92.15 \\
GPT-5.2 (think) / Wang       & 77 & 99.90 & 91.99 \\
GPT-5.2 (think) / Chudi\v{c} & 72 & 99.32 & 93.20 \\
GPT-5.2 (think) / Ou\'edraogo& 84 & 99.53 & 91.99 \\
\midrule
Gemini (low) / Ours          & 47 & 99.75 & 88.76 \\
Gemini (low) / Wang          & 67 & 99.82 & 93.67 \\
Gemini (low) / Chudi\v{c}    & 66 & 99.64 & 90.31 \\
Gemini (low) / Ou\'edraogo   & 68 & 99.55 & 91.49 \\
\midrule
Gemini (high) / Ours         & 52 & 99.94 & 88.31 \\
Gemini (high) / Wang         & 74 & 99.84 & 90.23 \\
Gemini (high) / Chudi\v{c}   & 71 & 99.67 & 89.32 \\
Gemini (high) / Ou\'edraogo  & 74 & 99.71 & 91.19 \\
\midrule
DeepSeek Chat / Ours         & 11 & 99.62 & 89.78 \\
DeepSeek Chat / Wang         & 20 & 99.68 & 89.17 \\
DeepSeek Chat / Chudi\v{c}   & 15 & 99.56 & 79.72 \\
DeepSeek Chat / Ou\'edraogo  & 25 & 99.72 & 87.50 \\
\midrule
DeepSeek Reason / Ours       &  7 & 99.76 & 85.00 \\
DeepSeek Reason / Wang       & 17 & 99.59 & 87.66 \\
DeepSeek Reason / Chudi\v{c} & 19 & 99.85 & 88.98 \\
DeepSeek Reason / Ou\'edraogo& 23 & 99.51 & 94.63 \\
\midrule
\midrule
{Claude Opus / Ours}            & 58 & 98.20 & 87.79 \\
{Claude Opus / Wang}            & 74 & 98.49 & 91.43 \\
{Claude Opus / Chudi\v{c}}      & 57 & 89.29 & 80.16 \\
{Claude Opus / Ou\'edraogo}     & 57 & 97.89 & 88.81 \\
\midrule
{Claude Opus (think) / Ours}    & 67 & 98.29 & 91.98 \\
{Claude Opus (think) / Wang}    & 74 & 98.34 & 88.35 \\
{Claude Opus (think) / Chudi\v{c}} & 67 & 95.24 & 84.24 \\
{Claude Opus (think) / Ou\'edraogo} & 67 & 96.45 & 85.39 \\
\midrule
{Claude Sonnet / Ours}          & 58 & 98.12 & 89.20 \\
{Claude Sonnet / Wang}          & 24 & 95.57 & 86.54 \\
{Claude Sonnet / Chudi\v{c}}    & 51 & 97.83 & 87.62 \\
{Claude Sonnet / Ou\'edraogo}   & 51 & 97.79 & 91.41 \\
\midrule
{Claude Sonnet (think) / Ours}  & 64 & 98.25 & 88.50 \\
{Claude Sonnet (think) / Wang}  & 42 & 97.40 & 89.83 \\
{Claude Sonnet (think) / Chudi\v{c}} & 55 & 98.04 & 86.78 \\
{Claude Sonnet (think) / Ou\'edraogo} & 59 & 98.09 & 89.19 \\
\bottomrule
\end{tabular}
\end{table*}

\subsection{Mutation Testing Results}\label{sec:mutation-results}

We report fault detection first within each model (holding the model fixed and varying the prompt) and then across models, using the mutation score from \texttt{CosmicRay}.

\begin{figure}[htbp]
\centering
\includegraphics[width=\textwidth]{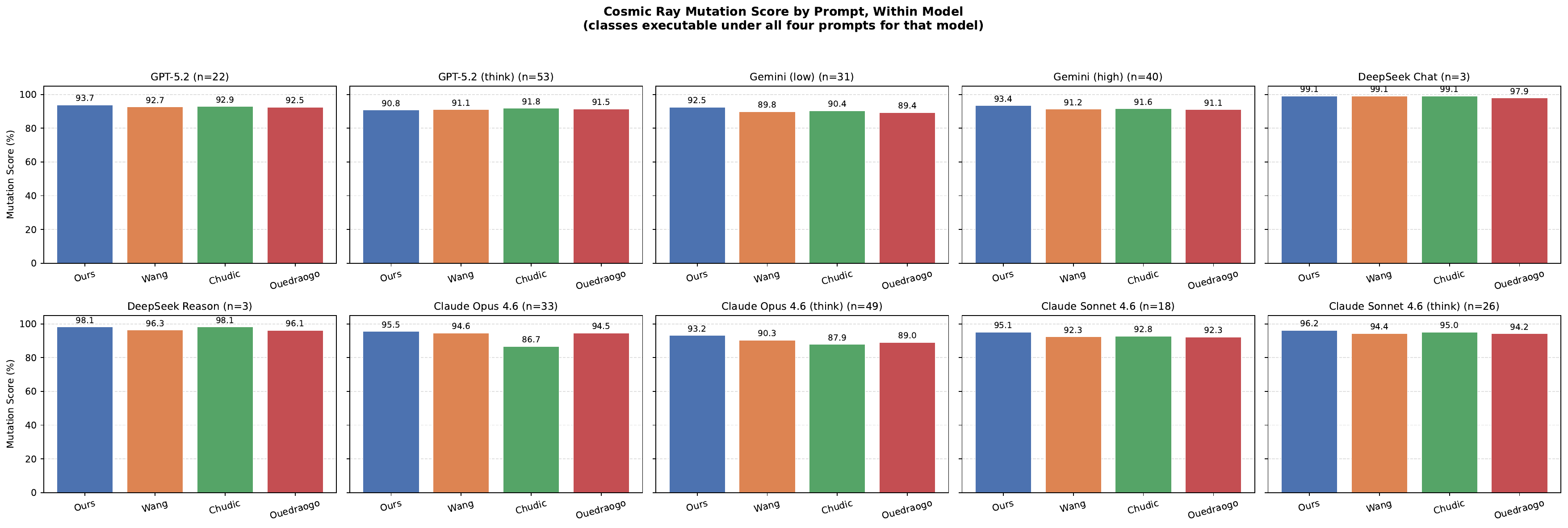}
\caption{Mutation Score (Cosmic Ray) (\%) by prompt, separately for each of the {ten} model configurations. Each panel is restricted to classes executable under the given (model, prompt) cell.}
\label{fig:model-across-prompts}
\end{figure}

\paragraph{Within a model} 
Within a model, mutation scores differ little once executability is controlled. \autoref{fig:model-across-prompts} compares the four prompts for each configuration. The proposed prompt leads in seven of ten configurations: GPT-5.2 without reasoning, both Gemini configurations, DeepSeek Reasoner, and three Claude configurations. Ouedraogo et al.'s prompt leads for GPT-5.2 with medium reasoning and for Claude Sonnet without thinking, the latter by 0.07 percentage points. The Chudic and \c{C}al{\i}kl{\i} prompt leads for DeepSeek Chat. As \autoref{tab:mutation-score-comparison} shows, the within-model range is usually below four percentage points; Claude Opus is the exception, with a range of up to 10 points.

Statistical significance is concentrated in the Claude cells. Of 703 eligible pairs in the 40-node dominance graph, 81 remain significant after Holm correction. Seventy-five significant edges involve at least one Claude cell, and a Claude cell is the winner in 72. The largest paired effect is a 7.7-point advantage for Claude Sonnet/Ours over Claude Opus (think)/Chudic ($n=50$, $p_{\mathrm{adj}}<0.001$). Fifty-two edges cross provider boundaries, while none involves a DeepSeek cell. The leading point estimates therefore remain statistically unresolved.

Reasoning effects differ by provider. \autoref{tab:thinking-effect} pairs each base configuration with its reasoning-enabled counterpart on shared executable classes. For GPT-5.2, medium reasoning raises executability from 46 to 78 classes. On the 41 paired classes, mutation score falls by 1.41 points ($p=0.031$) and assertions fall by 4.12 per suite ($p=0.0003$). For Gemini, executability rises from 47 to 52 classes; paired suites contain 2.83 more test functions and 6.83 more assertions, while mutation score rises by 0.64 points. Only four classes execute under both DeepSeek configurations, so that comparison is underpowered. For Claude, thinking raises executability for Opus (58 to 67) and Sonnet (58 to 64). It also reduces the size of paired Opus suites by 24.4 test functions and 33.4 assertions, with a 1.16-point reduction in mutation score. Sonnet's paired mutation score is essentially unchanged.

\begin{table*}[htbp]
\centering
\caption{Within-provider paired deltas (reasoning enabled minus reasoning disabled) on classes executable under both configurations, proposed prompt. Wilcoxon signed-rank $p$-values on the paired differences. Executable Test Percentage (\%) is reported on the full 100-class base. The DeepSeek row is flagged as underpowered ($n=4$).}
\label{tab:thinking-effect}
\footnotesize
\begin{tabular}{llrrrrr}
\toprule
\textbf{Provider} & \textbf{Metric} & \textbf{Off} & \textbf{On} & \textbf{Delta} & \textbf{n} & \textbf{p} \\
\midrule
OpenAI  & Mutation Score (Cosmic Ray) (\%) & 93.28 & 91.87 & $-1.41$ & 41  & 0.031 \\
OpenAI  & Executable Test Percentage (\%)  & 46.0  & 78.0  & $+32.0$ & 100 & n/a \\
OpenAI  & Number of Test Functions         & 15.51 & 14.41 & $-1.10$ & 41  & 0.065 \\
OpenAI  & Number of Assertions             & 24.93 & 20.80 & $-4.12$ & 41  & $0.0003$ \\
\midrule
Gemini  & Mutation Score (Cosmic Ray) (\%) & 92.34 & 92.99 & $+0.64$ & 36  & 0.023 \\
Gemini  & Executable Test Percentage (\%)  & 47.0  & 52.0  & $+5.0$  & 100 & n/a \\
Gemini  & Number of Test Functions         & 15.08 & 17.92 & $+2.83$ & 36  & $5.7\mathrm{e}{-5}$ \\
Gemini  & Number of Assertions             & 23.72 & 30.56 & $+6.83$ & 36  & $7.1\mathrm{e}{-6}$ \\
\midrule
DeepSeek$^{\dagger}$ & Mutation Score (Cosmic Ray) (\%) & 96.57 & 97.59 & $+1.02$ & 4   & 1.000 \\
DeepSeek$^{\dagger}$ & Executable Test Percentage (\%)  & 11.0  & 7.0   & $-4.0$  & 100 & n/a \\
DeepSeek$^{\dagger}$ & Number of Test Functions         & 33.75 & 33.25 & $-0.50$ & 4   & 1.000 \\
DeepSeek$^{\dagger}$ & Number of Assertions             & 55.25 & 60.00 & $+4.75$ & 4   & 0.625 \\
\midrule
Claude Opus   & Mutation Score (Cosmic Ray) (\%) & 94.31 & 93.16 & $-1.16$  & 52  & $0.0022$ \\
Claude Opus   & Executable Test Percentage (\%)  & 58.0  & 67.0  & $+9.0$   & 100 & n/a \\
Claude Opus   & Number of Test Functions         & 56.17 & 31.79 & $-24.38$ & 52  & $3.5\mathrm{e}{-10}$ \\
Claude Opus   & Number of Assertions             & 73.48 & 40.06 & $-33.42$ & 52  & $3.5\mathrm{e}{-10}$ \\
\midrule
Claude Sonnet & Mutation Score (Cosmic Ray) (\%) & 94.55 & 94.45 & $-0.10$  & 51  & $0.744$ \\
Claude Sonnet & Executable Test Percentage (\%)  & 58.0  & 64.0  & $+6.0$   & 100 & n/a \\
Claude Sonnet & Number of Test Functions         & 55.43 & 51.73 & $-3.71$  & 51  & $0.0045$ \\
Claude Sonnet & Number of Assertions             & 66.63 & 60.39 & $-6.24$  & 51  & $0.0010$ \\
\bottomrule
\end{tabular}

\vspace{2pt}
\footnotesize $^{\dagger}$ Underpowered: only four classes executable under both DeepSeek configurations.
\end{table*}

\subsubsection*{Across-Model Comparisons}
\autoref{tab:mutation-score-comparison} reports the ten-by-four grid. Averaged across prompts, DeepSeek Chat has the highest mean on its own executable subsets (94.6\%), followed by Claude Sonnet and DeepSeek Reasoner (both 94.1\%). The remaining model means range from 89.9\% to 93.8\%. These rankings are descriptive, not like-for-like: DeepSeek means use only 7 to 25 classes per cell, whereas other cells include as many as 84. DeepSeek's high conditional mutation scores must therefore be read alongside its low executability.

\begin{table}[htbp]
\footnotesize
\centering
\caption{Mutation Score (Cosmic Ray) (\%) by model and prompt, restricted to executable classes. Bold marks the best prompt per model. Cell counts show the size of the executable subset. {For Claude Sonnet (no thinking), the best prompt is Ou\'edraogo at 94.53\% against Ours at 94.46\%; both round to 94.5.}}
\label{tab:mutation-score-comparison}
\begin{tabular}{lcccc}
\toprule
Model & Ours & Wang & Chudi\v{c} & Ou\'edraogo \\
\midrule
GPT-5.2              & \textbf{93.3 ($n{=}46$)} & 90.7 ($n{=}52$) & 91.5 ($n{=}46$) & 92.5 ($n{=}63$) \\
GPT-5.2 (think)      & 90.2 ($n{=}78$) & 90.7 ($n{=}77$) & 90.8 ($n{=}72$) & \textbf{91.2 ($n{=}84$)} \\
Gemini (low)         & \textbf{91.8 ($n{=}47$)} & 90.8 ($n{=}67$) & 91.7 ($n{=}66$) & 89.5 ($n{=}68$) \\
Gemini (high)        & \textbf{93.4 ($n{=}52$)} & 91.6 ($n{=}74$) & 91.5 ($n{=}71$) & 91.3 ($n{=}74$) \\
DeepSeek Chat        & 94.8 ($n{=}11$) & 94.5 ($n{=}20$) & \textbf{95.0 ($n{=}15$)} & 94.0 ($n{=}25$) \\
DeepSeek Reason      & \textbf{96.2 ($n{=}7$)}  & 92.2 ($n{=}17$) & 94.1 ($n{=}19$) & 93.9 ($n{=}23$) \\
Claude Opus         & \textbf{94.3 ($n{=}58$)} & 93.6 ($n{=}74$) & 84.3 ($n{=}57$) & 94.0 ($n{=}57$) \\
Claude Opus (think) & \textbf{93.1 ($n{=}67$)} & 89.6 ($n{=}74$) & 88.7 ($n{=}67$) & 88.1 ($n{=}67$) \\
Claude Sonnet       & 94.5 ($n{=}58$) & 93.4 ($n{=}24$) & 94.1 ($n{=}51$) & \textbf{94.5 ($n{=}51$)} \\
Claude Sonnet (think)& \textbf{94.7 ($n{=}64$)} & 92.6 ($n{=}42$) & 94.6 ($n{=}55$) & 93.4 ($n{=}59$) \\
\bottomrule
\end{tabular}
\end{table}

\begin{figure}[htbp]
\centering
\includegraphics[width=\textwidth]{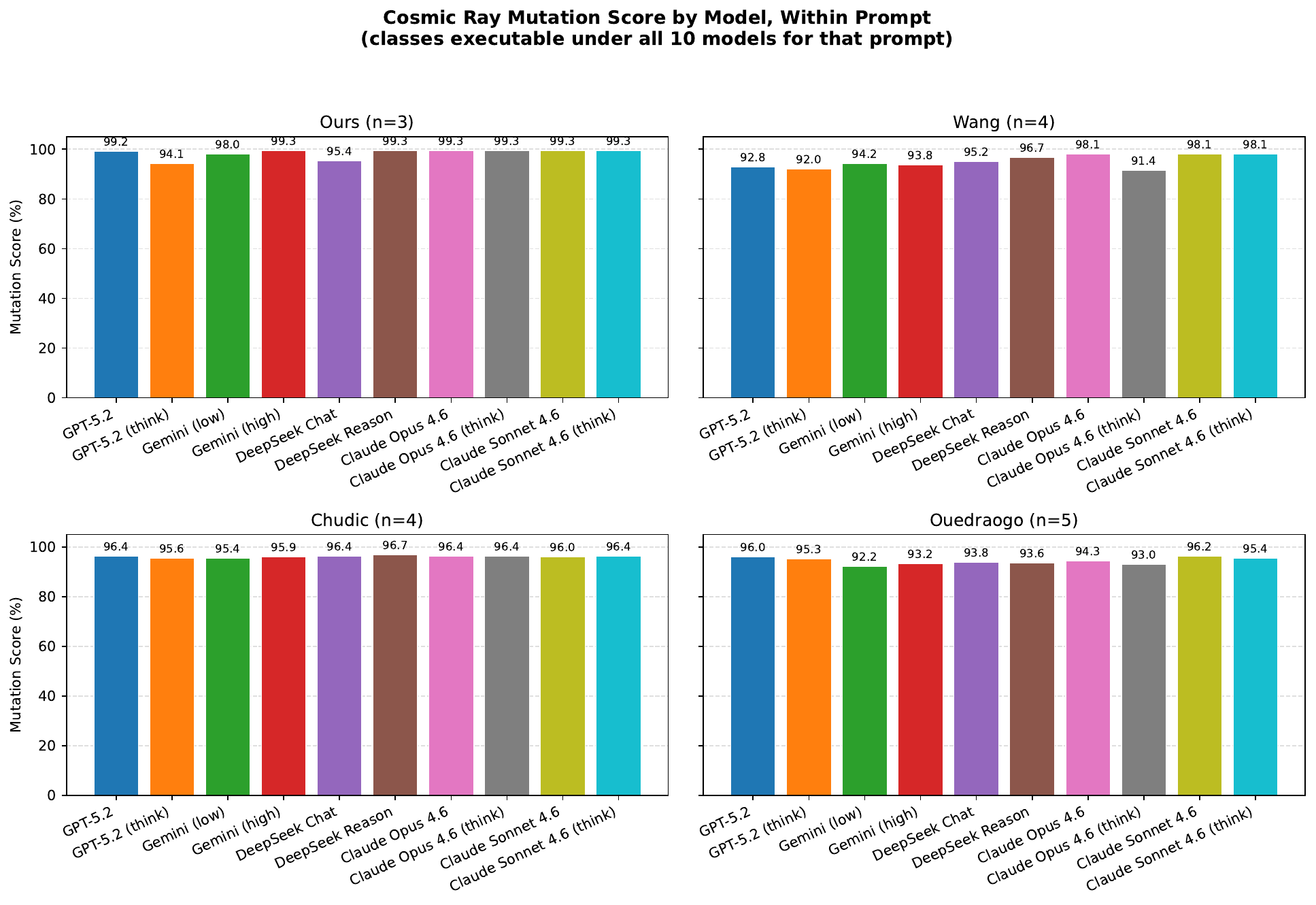}
\caption{Mutation Score (Cosmic Ray) (\%) by model, separately for each of the four prompts. Panels are restricted to executable classes. {(Regenerated from the values in \autoref{tab:mutation-score-comparison}.)}}
\label{fig:prompt-across-models}
\end{figure}

\begin{figure}[htbp]
\centering
\includegraphics[width=0.85\textwidth]{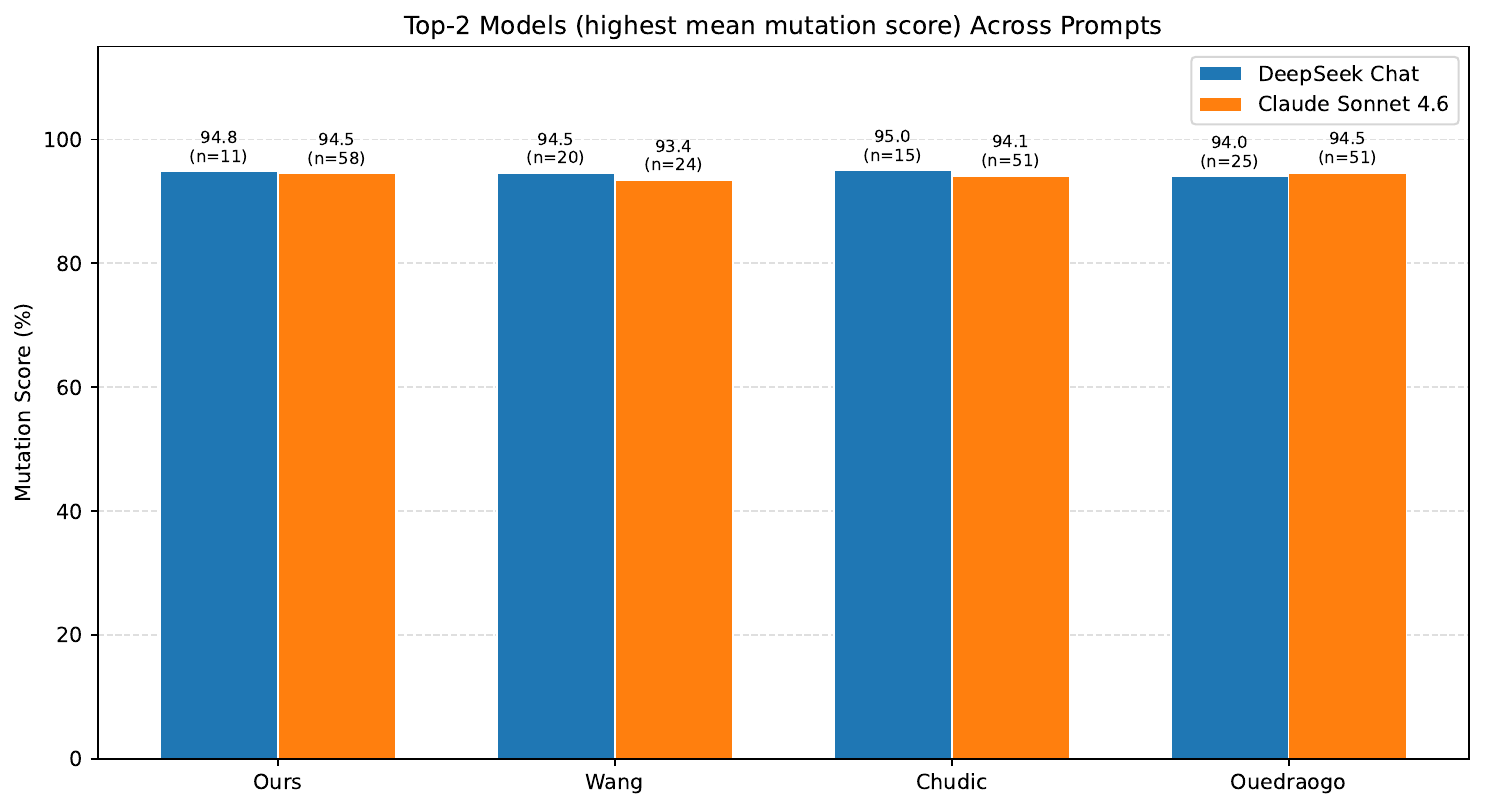}
\caption{Mutation Score (Cosmic Ray) (\%) for the {two models with the highest overall means (DeepSeek Chat and Claude Sonnet)} across the four prompts. {Cell sizes range from 11 to 58 classes}; error bars indicate standard error.}
\label{fig:top2-models}
\end{figure}

\begin{figure}[htbp]
\centering
{\includegraphics[width=\textwidth]{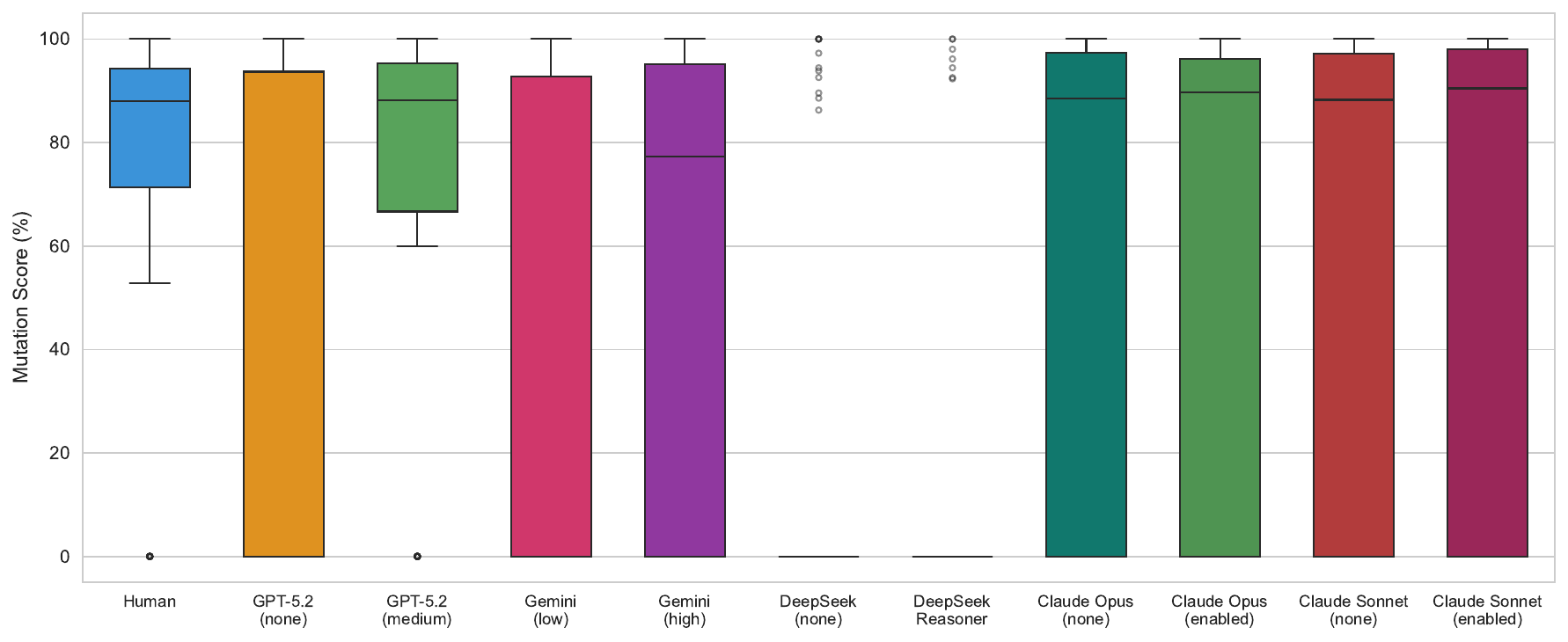}}
\caption{{Per-suite mutation-score distributions by (model, prompt) cell on executable classes. The medians cluster tightly (roughly {90--98\%}), consistent with the small main effect of prompt reported in \autoref{tab:mutation-score-comparison}.}}
\label{fig:mutation-box}
\end{figure}

\autoref{fig:prompt-across-models} rotates the same data along the prompt axis. Under the proposed prompt, DeepSeek Reasoner and DeepSeek Chat occupy the top two positions; under Wang, Chudic, and Ouedraogo a Claude cell takes one of the top two slots (Claude Opus, Claude Sonnet with thinking, and Claude Sonnet, respectively) with DeepSeek Chat holding the other; within every prompt the GPT and Gemini cells cluster within roughly three percentage points of one another. \autoref{fig:top2-models} zooms into the eight cells of {the two models with the highest overall means (DeepSeek Chat and Claude Sonnet)} and makes the tight clustering of the executable-subset means visible. \autoref{fig:mutation-box} shows the corresponding per-suite mutation-score distributions.

The Type-II two-way analysis of variance uses 2{,}099 executable rows. The model main effect is $\eta^2=0.0216$ ($F(9,2059)=5.21$, $p<0.001$), and the prompt main effect is $\eta^2=0.0056$ ($F(3,2059)=4.03$, $p=0.007$). Their ratio is 3.9. The interaction is also significant ($\eta^2=0.0238$, $F(27,2059)=1.91$, $p=0.003$), reflecting stronger prompt sensitivity among Claude configurations. For example, Claude Opus without thinking varies by 10.0 points across prompts, while every non-Claude model varies by no more than four points. Model choice has the larger main effect, but prompt choice and the model--prompt interaction are both statistically detectable.

\subsection{Test Quality Metrics}\label{sec:test-quality}

Suite volume varies sharply by model. \autoref{tab:summary-stats} reports the proposed-prompt results on executable classes. Human suites contain 22.85 test functions and 34.71 assertions on average. GPT-5.2 produces 13.77 to 15.54 functions and 20.31 to 25.43 assertions, depending on reasoning. DeepSeek produces about 32 to 33 functions and 56 to 62 assertions. Mean suite size ranges from 116 to 151 lines for GPT-5.2 and Gemini, 285 to 294 for DeepSeek, and 177 to 376 for Claude; the human mean is 148 lines. These large volume differences correspond to less than five points of mutation-score variation across model means. Larger suites do not necessarily detect more faults.

\begin{table*}[htbp]
\centering
\caption{Summary statistics for the proposed prompt, restricted to executable classes. Values are mean (SD). LLM rows use the proposed prompt only; the human row uses the 82 ClassEval classes whose human-written suite executed.}
\label{tab:summary-stats}
\resizebox{\textwidth}{!}{%
\begin{tabular}{lrrrrrr r rrrr}
\toprule
\textbf{Metric} & \textbf{Human} & \textbf{GPT-5.2} & \textbf{GPT-5.2 (think)} & \textbf{Gemini (low)} & \textbf{Gemini (high)} & \textbf{DeepSeek Chat} & \textbf{DeepSeek Reason} & \textbf{Claude Opus} & \textbf{Claude Opus (think)} & \textbf{Claude Sonnet} & \textbf{Claude Sonnet (think)} \\
\midrule
Mutation Score (Cosmic Ray) (\%)      & 88.26 (10.77) & 93.31 (7.63)  & 90.17 (9.09)  & 91.81 (8.31)  & 93.38 (7.13)  & 94.78 (5.09)  & 96.21 (3.26)  & 94.30 (6.62) & 93.13 (6.85) & 94.46 (6.17) & 94.69 (5.77) \\
Executable Test Percentage (\%)        & 82.0          & 46.0          & 78.0          & 47.0          & 52.0          & 11.0          & 7.0           & 58.0 & 67.0 & 58.0 & 64.0 \\
Number of Test Functions               & 22.85 (7.95)  & 15.54 (3.67)  & 13.77 (3.89)  & 15.06 (3.01)  & 18.04 (5.55)  & 33.18 (6.65)  & 32.29 (4.92)  & 56.14 (13.27) & 32.70 (9.27) & 55.69 (13.94) & 52.73 (13.90) \\
Number of Assertions                   & 34.71 (18.47) & 25.43 (9.32)  & 20.31 (9.37)  & 23.91 (8.07)  & 29.62 (11.47) & 55.64 (16.56) & 62.29 (15.45) & 73.03 (17.37) & 42.24 (13.10) & 66.50 (16.92) & 61.44 (17.34) \\
Assertion Density                      & 1.53 (0.72)   & 1.69 (0.65)   & 1.50 (0.64)   & 1.57 (0.37)   & 1.63 (0.33)   & 1.70 (0.45)   & 1.97 (0.59)   & 1.31 (0.22) & 1.30 (0.25) & 1.20 (0.15) & 1.16 (0.11) \\
Mean Cyclomatic Complexity per Test    & 1.09 (0.32)   & 2.60 (0.70)   & 2.34 (0.69)   & 2.53 (0.41)   & 2.59 (0.42)   & 2.81 (0.60)   & 2.98 (0.59)   & 2.34 (0.27) & 2.32 (0.28) & 2.25 (0.19) & 2.22 (0.16) \\
Test Suite Size (LOC)                  & 148.05 (78.74)& 150.87 (30.06)& 141.63 (48.02)& 116.13 (16.86)& 143.00 (30.73)& 284.91 (50.19)& 294.00 (50.77)& 337.67 (67.11) & 176.84 (46.18) & 376.41 (68.82) & 318.88 (72.26) \\
Test-to-Source LOC Ratio               & 3.94 (2.14)   & 3.93 (1.32)   & 3.73 (1.33)   & 3.24 (0.87)   & 4.04 (1.04)   & 8.41 (3.82)   & 8.52 (2.99)   & 9.34 (2.92) & 4.58 (1.32) & 10.76 (3.90) & 8.64 (2.99) \\
\bottomrule
\end{tabular}}
\end{table*}

\begin{figure}[htbp]
\centering
\includegraphics[width=\textwidth]{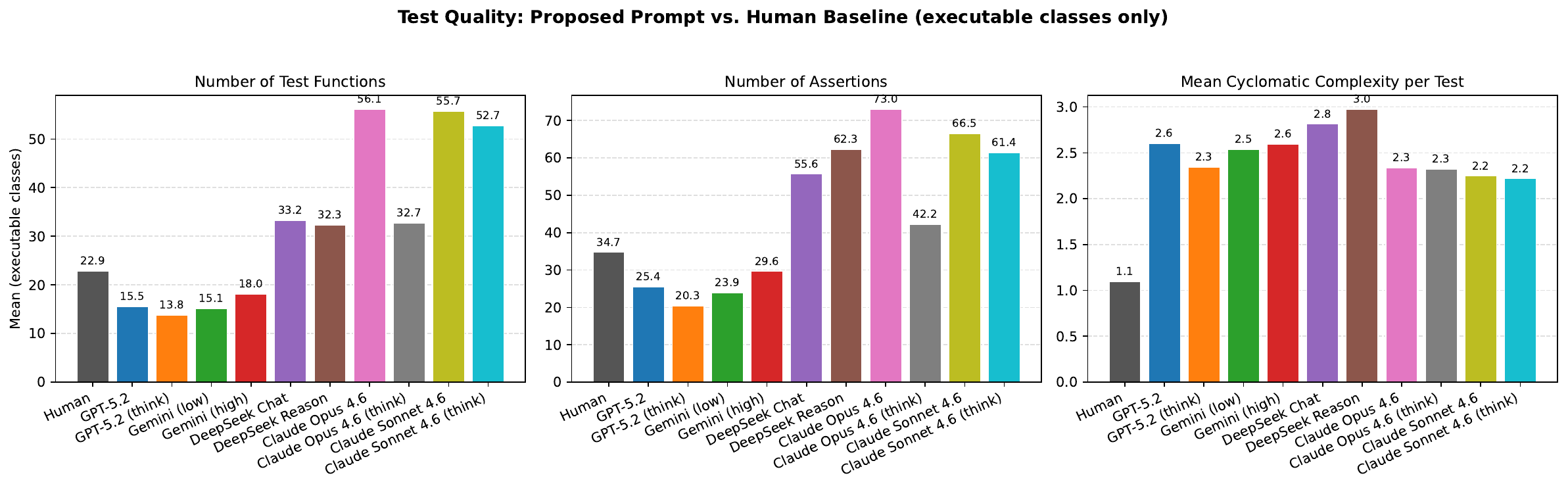}
\caption{Test quality panel: number of test functions (left), number of assertions (middle), and mean cyclomatic complexity per test (right). All panels restricted to executable classes. LLMs configurations cluster far above humans on complexity regardless of provider or reasoning level.}
\label{fig:test-quality-panel}
\end{figure}

\begin{figure}[htbp]
\centering
{\includegraphics[width=\textwidth]{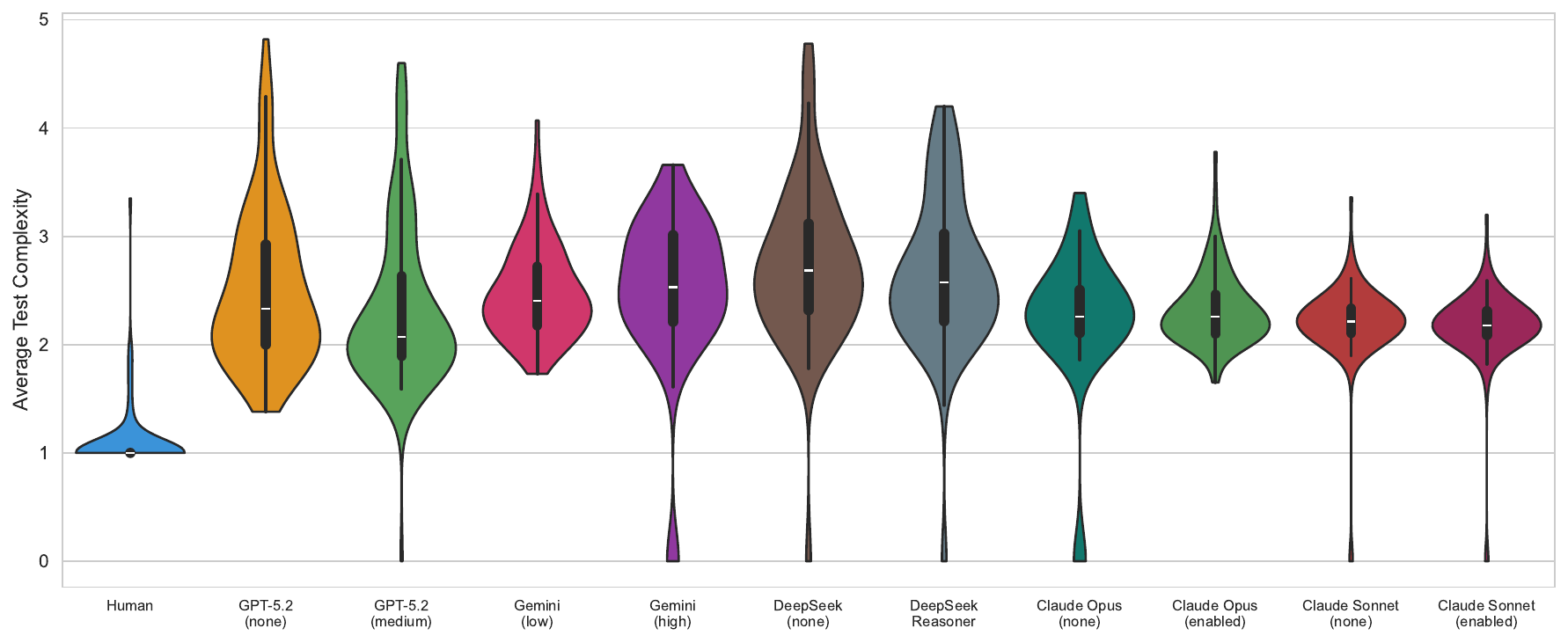}}
\caption{{Distribution of mean cyclomatic complexity per test by configuration. Every LLM configuration sits two to three times above the human baseline (1.09), a uniform structural signature of LLM-generated tests.}}
\label{fig:complexity-violins}
\end{figure}

\autoref{fig:test-quality-panel} shows suite-level distributions for test functions, assertions, and mean cyclomatic complexity per test. The first two panels separate high-volume DeepSeek and Claude suites from GPT-5.2, Gemini, and the human baseline. Complexity provides a clearer distinction: LLM means range from 2.22 to 2.98, compared with 1.09 for human-written tests. This descriptive gap appears across providers and reasoning levels. \autoref{fig:complexity-violins} displays the complexity distributions separately.

\subsection{Correlation Between Evaluation Metrics}\label{sec:correlation}

\begin{figure}[htbp]
\centering
{\includegraphics[width=0.9\textwidth]{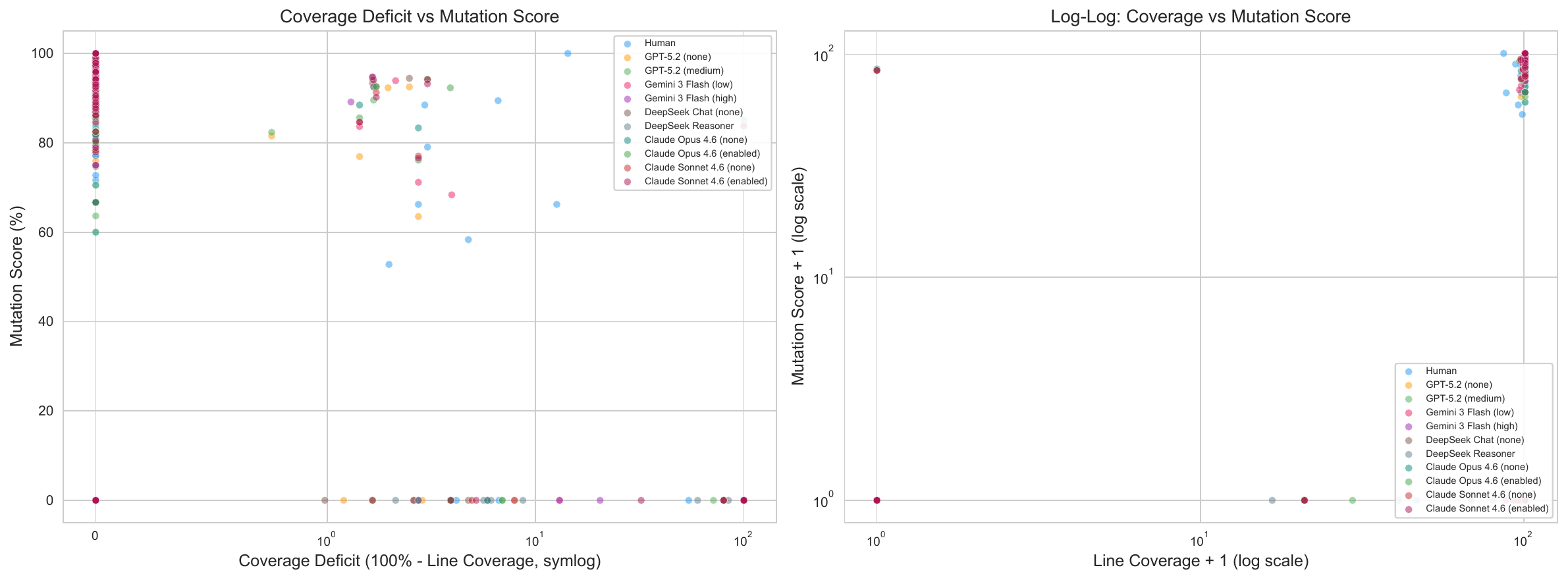}}
\caption{{Mutation score versus line coverage on executable classes. Coverage is saturated near 100\% while mutation score varies, illustrating the weak coverage--fault-detection relationship.}}
\label{fig:cov-vs-mut}
\end{figure}

\begin{figure}[htbp]
\centering
\includegraphics[width=\textwidth]{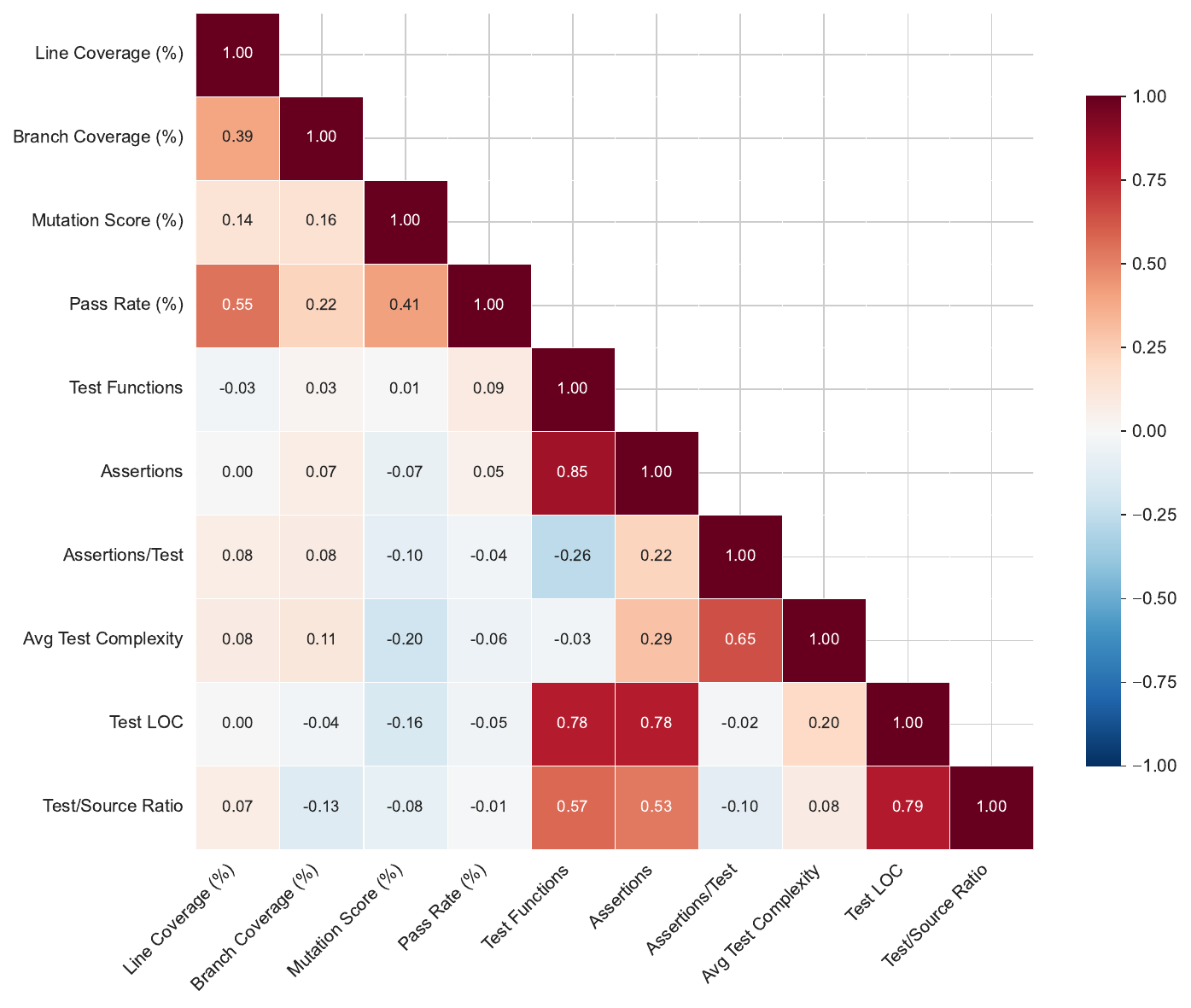}
\caption{Pearson correlation heatmap of the evaluation metrics. Mutation score correlates only weakly with line/branch coverage ($r=0.11/0.10$) and negatively with the volume metrics (test functions, assertions, LOC), while the volume metrics correlate strongly with one another.}
\label{fig:correlation-heatmap}
\end{figure}

Next, we examine relationships among the metrics. \autoref{fig:cov-vs-mut} plots mutation score against line coverage on executable suites. Coverage is compressed near its ceiling while mutation score varies, so the relationship is weak. \autoref{fig:correlation-heatmap} shows correlations of $r=0.11$ with line coverage and $r=0.10$ with branch coverage. Mutation score correlates moderately with pass rate ($r=0.40$) and negatively with suite-volume measures: test functions ($r=-0.31$), assertions ($r=-0.27$), test lines ($r=-0.39$), and test-to-source ratio ($r=-0.33$). In contrast, the volume measures correlate strongly with one another, including $r=0.80$ between assertions and test functions. Suite size and fault detection therefore capture different properties.

\subsection{Comparative Results Across LLMs and Prompts}\label{sec:cross-prompt-synthesis}

No prompt dominates every configuration. The proposed prompt has the highest conditional mean in seven configurations, Ouedraogo et al.'s prompt in two, and the Chudic and \c{C}al{\i}kl{\i} prompt in one. The executable-gated ANOVA estimates a small but significant prompt main effect ($\eta^2=0.0056$, $p=0.007$) and a model effect 3.9 times larger. In the zero-filled Friedman sensitivity analysis, four models reject equal prompt effects: DeepSeek Reasoner and three Claude configurations. Because that sensitivity analysis assigns zero to failed suites, its result reflects both executability and mutation score. \autoref{sec:discussion} interprets these findings and the dominance diagrams.

\section{Discussion}\label{sec:discussion}

This section interprets the results. We first answer the research questions (\autoref{sec:rq-synthesis}), then compare our findings with related approaches (\autoref{sec:comparison}), distil the key findings and their implications including the dominance diagrams (\autoref{sec:key-findings}), and finally assess threats to validity (\autoref{sec:threats}).

\subsection{Discussion of Research Questions}\label{sec:rq-synthesis}

\subsubsection*{RQ1: Does prompt strategy significantly affect structural coverage?}
We find no practically useful prompt effect after conditioning on executability. Median line and branch coverage is 100\% in every model--prompt cell. Mean line coverage ranges from 99.3\% to 99.95\% outside Claude and reaches a minimum of 89.3\% within Claude cells (\autoref{tab:app-coverage}). Coverage-based rankings therefore cannot separate prompt strategies reliably in this benchmark.

\subsubsection*{RQ2: Does prompt strategy significantly affect fault detection?}
Yes, but the effect is small and model-dependent. On executable rows, the prompt main effect is $\eta^2=0.0056$ ($F(3,2059)=4.03$, $p=0.007$), roughly one quarter of the model effect. Eighty-one of 703 eligible pairwise comparisons remain significant after Holm correction, and 75 involve Claude cells. The proposed prompt has the highest mean in seven configurations, but within-model ranges usually remain below four percentage points. Claude Opus is the exception, with ranges up to 10 points. Prompt strategy therefore affects fault detection, but no prompt is uniformly best.

\subsubsection*{RQ3: What is the relative importance of model selection versus prompt design?}
Model selection dominates. The model main effect on mutation score ($\eta^2=0.022$) is 3.9 times the prompt main effect ($\eta^2=0.006$). DeepSeek Chat and DeepSeek Reasoner lead three of the four prompts on mutation score at 94.55\% and 94.09\% mean (Claude Sonnet, at 94.12\%, takes the fourth), but they execute on only 17.8\% and 16.5\% of the 100 ClassEval classes; GPT-5.2 with medium reasoning trails on mean mutation score (90.7\%) but executes on 77.8\%. The binding constraint shifts by model: DeepSeek is limited by executability, GPT-5.2 and Gemini are limited by assertion strength. We conclude that model selection dominates prompt design, and the two are complementary levers: prompts move a model by a few percentage points on mutation score, whereas models differ in the proportion of classes for which they produce executable tests by tens of percentage points.

\subsection{Comparison with Related Approaches}\label{sec:comparison}

Our primary contribution relative to prior work is the evaluation framework itself, not any single performance number. No previous study of LLM-generated tests reports mutation scores for class-level Python code alongside structural coverage, Executable Test Percentage, and quality metrics within a controlled multi-prompt, multi-model design. We position our results against the existing literature on two dimensions: coverage (the metric we share with prior work) and mutation score (the metric we introduce to this domain).

On executable suites, median line coverage reaches 100\% in every cell; mean line coverage ranges from 89.3\% to 99.95\%. These values exceed some reported results for ChatUniTest and Code Llama~\citep{chen2024chatunitest} and for GPT-4o on seven code blocks~\citep{diazarrieta2025comparative}. Other benchmarks report line coverage from roughly 38\% to above 80\%~\citep{wang2025projecttest,pan2025aster,wang2025testeval}. Direct numerical comparison is limited because the benchmarks, languages, and executable filters differ~\citep{wang2025projecttest,pan2025aster,chen2024chatunitest,wang2025testeval,diazarrieta2025comparative,alves2024detecting,yang2025rted}. Within our factorial design, coverage cannot rank the configurations reliably.

Fault detection requires a more cautious comparison. Many studies emphasize coverage~\citep{wang2025projecttest,wang2025testeval}, while our protocol separates executability from conditional mutation score. Human-written suites execute for 82 classes and kill 88.3\% of mutants on that subset. LLM configuration means range from 90.2\% to 96.2\%, but each uses its own executable subset. These unequal denominators prevent a direct claim that LLM-generated tests outperform human-written tests. DeepSeek Chat illustrates the trade-off: its mean is 94.8\% on 11 executable classes, but the other 89 classes fail the gate. Reporting both quantities avoids treating selective execution as general superiority.

\subsection{Key Findings and Implications}\label{sec:key-findings}

Five findings emerge from the results; we interpret each here rather than restating the underlying numbers.

\paragraph{Multi-Dimensional Evaluation Is Necessary.} 
Coverage alone produces a misleading account of LLM-generated tests. DeepSeek cells approach 100\% mean line coverage on the suites that execute, yet fail the executability gate for 75\% to 93\% of classes. GPT-5.2 with medium reasoning executes far more often. This result extends the established finding that coverage is weakly related to test effectiveness~\citep{inozemtseva2014coverage} and adds a distinct failure mode: some generated suites never run. Evaluation should therefore report executability and mutation score alongside coverage.

\paragraph{Executability and Mutation Score Diverge.} Prompts differ more in how often their output runs than in conditional mutation score. Ouedraogo et al.'s prompt clears the gate in 57.1\% of model--class cells, compared with 48.8\% for the proposed prompt. Among executable suites, within-model mutation-score ranges are usually below four points. Prompt design should therefore prioritize valid imports and execution before pursuing marginal gains in conditional mutation score.

\paragraph{Model Selection Dominates Prompt Choice} Model and reasoning configuration account for 3.9 times as much mutation-score variance as prompt choice ($\eta^2=0.0216$ versus $0.0056$). This does not make prompts irrelevant: the significant interaction shows that some models are more prompt-sensitive than others. In practice, teams must balance conditional fault detection against the probability that a generated suite will execute.

\paragraph{Reasoning Effects Are Asymmetric} Enabling reasoning helps differently by provider. For GPT-5.2, it raises executability from 46\% to 78\% while reducing assertions on paired suites. For Gemini, it increases test functions and assertions with only a small mutation-score change. The DeepSeek comparison is underpowered. Reasoning is therefore a model-specific control, not a uniformly beneficial setting.

\begin{figure}[htbp]
\centering
\includegraphics[width=\textwidth]{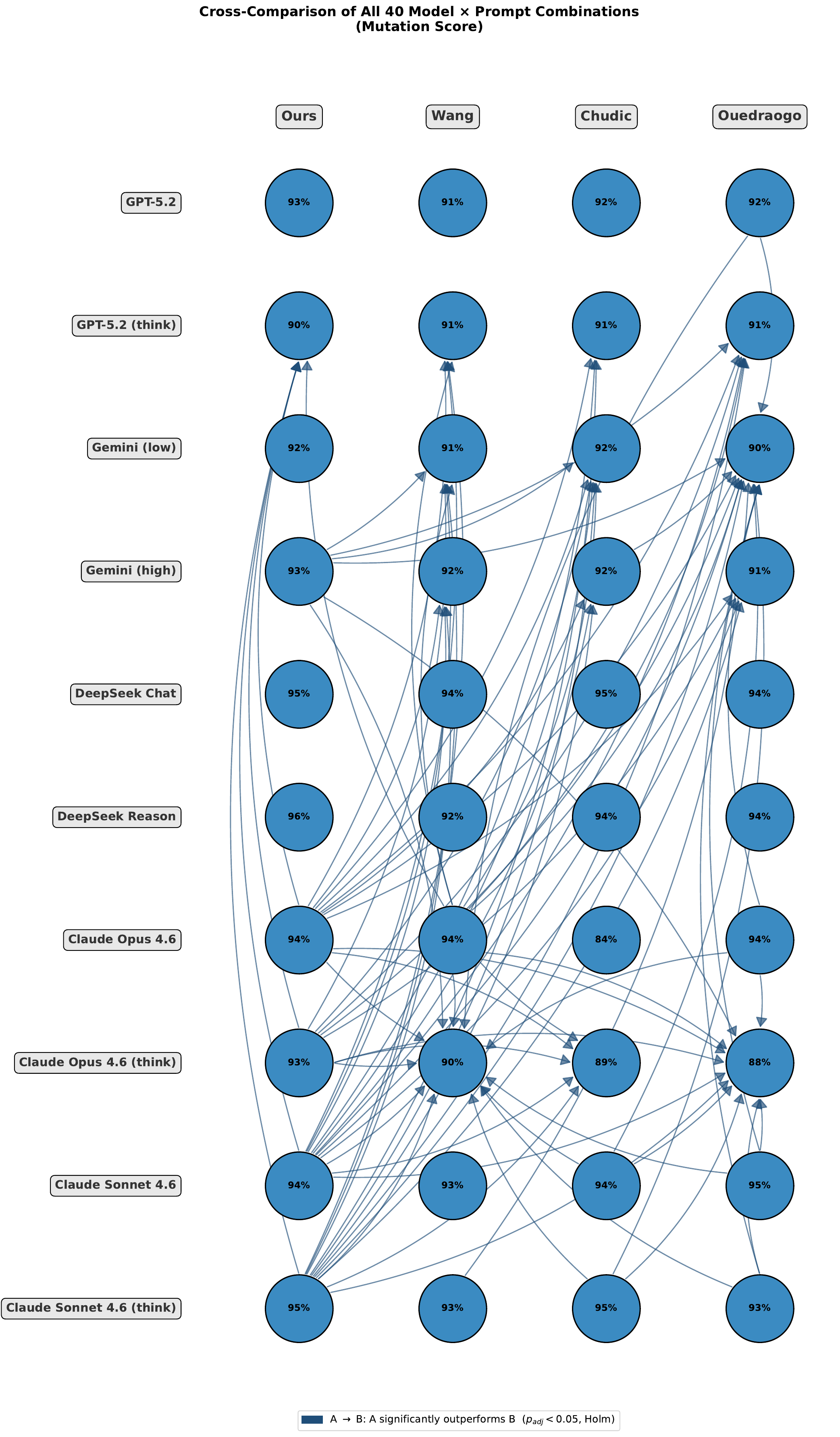}
\caption{Dominance diagram (cross graph) over all 40 (model, prompt) cells. A directed edge $A\rightarrow B$ indicates that $A$ significantly outperforms $B$ on mutation score under a Holm-corrected Wilcoxon signed-rank test ($p_{\mathrm{adj}}<0.05$). Eighty-one of 703 eligible pairs survive correction.}
\label{fig:mutation-cross-all24}
\end{figure}

\begin{figure}[htbp]
\centering
\includegraphics[width=0.85\textwidth]{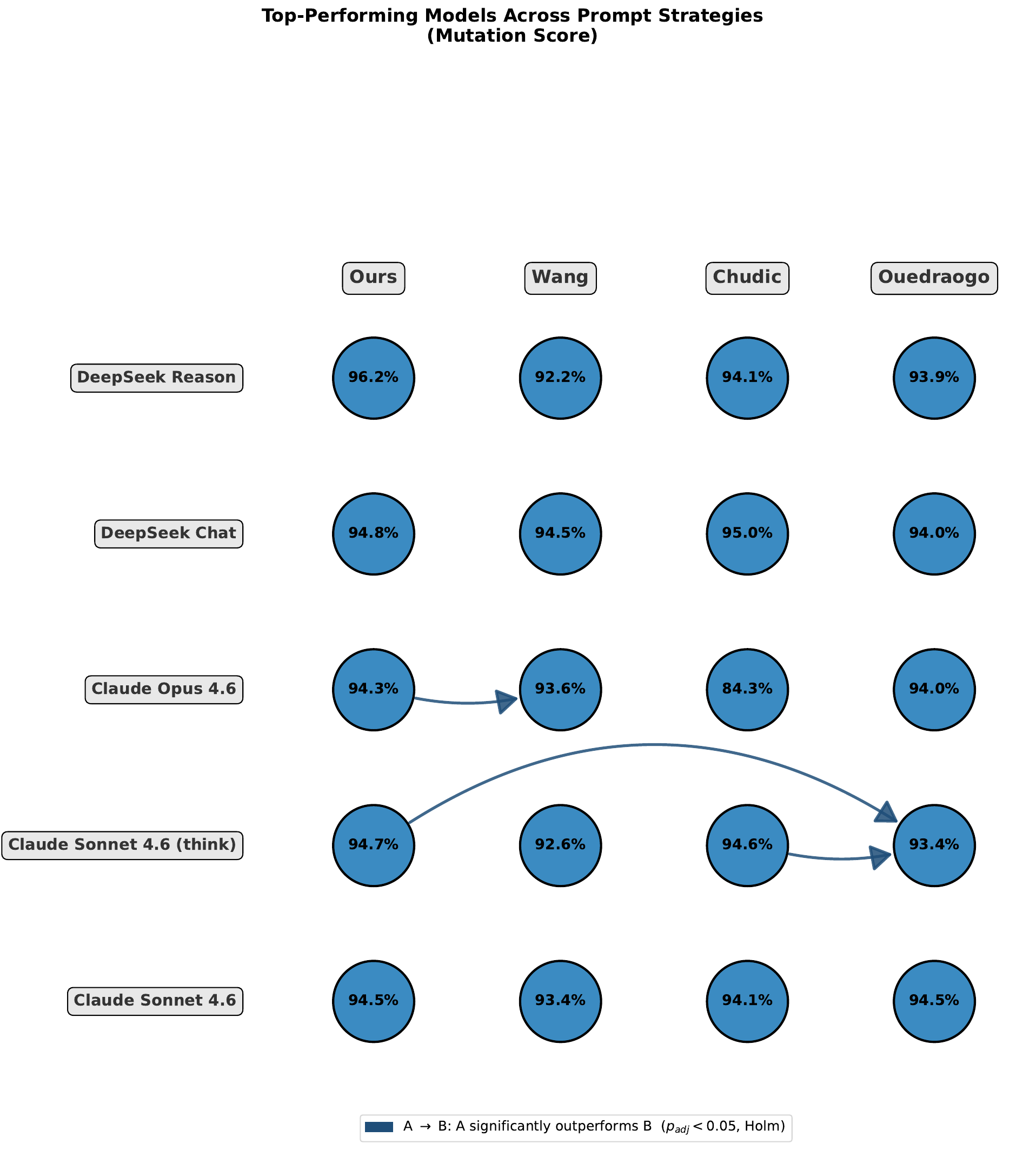}
\caption{Dominance diagram for five high-scoring model configurations: DeepSeek Chat, DeepSeek Reasoner, Claude Opus, Claude Sonnet, and Claude Sonnet with thinking. Three within-model prompt edges survive Holm correction; no edge separates different models in this restricted set.}
\label{fig:mutation-cross-topmodels}
\end{figure}

\paragraph{Dominance Among Configurations.} To summarize which configurations significantly outperform which, we use \emph{dominance diagrams}; see \autoref{fig:mutation-cross-all24} and \autoref{fig:mutation-cross-topmodels}. 
A dominance diagram represents each model--prompt configuration as a node. An edge $A \rightarrow B$ means that $A$ has a higher mutation score under a Holm-corrected Wilcoxon test ($p_{\mathrm{adj}}<0.05$); no edge means that the analysis detects no significant difference. In the full 40-cell graph, 81 of 703 eligible comparisons survive correction. Seventy-five involve a Claude cell, and none involves DeepSeek. The restricted graph contains three within-model edges: Claude Opus/Ours over Claude Opus/Wang, and Claude Sonnet (think)/Ours and /Chudic over /Ouedraogo. No edge separates one model from another in this restricted set. DeepSeek Reasoner/Ours has the largest point estimate (96.2\%, $n=7$), but its small executable subset limits pairwise power.

\paragraph{Implications for practitioners.} 
Based on our research, we have four major recommendations for practioners of software engineering. First, report executability as a primary outcome as LLM-generated test cases may not even be executable in many cases. 
Second, select the model and reasoning setting before fine-tuning prompts as they often matter more than prompts themselves. 
Third, combine mutation testing with an explicit execution gate. 
Finally, do not equate suite volume with quality: larger DeepSeek suites execute for fewer classes than GPT-5.2 suites and do not deliver a commensurate increase in conditional mutation score.

\subsection{Threats to Validity}\label{sec:threats}

Four caveats bound our claims. First, we evaluate one generation per configuration on a single benchmark (ClassEval, 100 Python classes), so the results speak to class-level Python code; other languages and project-level settings are natural extensions. Second, Cosmic Ray's operators cover the standard syntactic mutation classes, and equivalent mutants depress absolute scores uniformly across configurations, therefore relative comparisons are unaffected. However, the choice of mutation tool itself conditions our results. Third, executable-subset sizes vary by cell (7 to 84 classes), which limits the power of some pairwise contrasts; we mitigate this with Holm-corrected paired tests, a minimum-$n$ eligibility rule, and small-$n$ flags in \autoref{tab:thinking-effect}.
Finally, the LLM services and model versions are time-specific. Provider updates may limit exact replication even when model identifiers and prompts are preserved.

\section{Conclusion}\label{sec:conclusion}

\subsection{Summary}\label{sec:summary}

This study separates prompt effects from model effects in class-level Python test generation. We evaluate four prompts across ten LLM configurations and 100 ClassEval classes with an execution gate, structural coverage, mutation testing, and quality indicators. Coverage reaches or approaches its ceiling and cannot rank the configurations. Executability varies from 7\% to 84\%, while conditional mutation-score means range from 90.2\% to 96.2\%. The proposed prompt leads in seven configurations but does not dominate every model. Model choice has a main effect 3.9 times larger than prompt choice, and the significant interaction shows that prompt sensitivity depends on the model. Human-written suites execute more often than most LLM-generated suites. Their mutation score and the LLM scores use different executable subsets, so they should not be interpreted as a direct contest. The central practical result is straightforward: choose models for reliable execution, then tune prompts, and evaluate successful suites with mutation testing rather than coverage alone. To our knowledge this is the first study to compare multiple prompt strategies across multiple LLMs using mutation testing on class-level Python code under an explicit executability gate.

\subsection{Future Work}\label{sec:future-work}

Future work should repeat generations under varied sampling settings to quantify stochasticity. Evaluation across more projects, languages, and task granularities would test generalizability. Methodological extensions include additional prompt strategies, stronger output repair, and more scalable mutation analysis. A direct comparison with mutation-guided generators such as MutGen, MuTAP, and Meta ACH would also show how much iterative mutation feedback adds beyond static prompting~\citep{wang2025mutgen,mutap,metaach2026}. Such a comparison should preserve class-level Python scope or state clearly which differences arise from language and granularity.

\backmatter

\section*{Acknowledgements}
This research used the High-Performance Computing (HPC) environment at the
American University of Sharjah (AUS). The authors thank the AUS-HPC team for the
computational infrastructure and technical support.

\section*{Declarations}

\subsection*{Funding}
This work was supported in part by the American University of Sharjah through
M.~Harshvardhan's Faculty Start-up Grant (SB2602). The funder had no role in
the study design, the collection and analysis of data, the decision to
publish, or the preparation of the manuscript.

\subsection*{Conflict of Interest}
The authors declare that they have no competing financial or non-financial
interests relevant to the content of this article.

\subsection*{Ethical Approval.}
Not applicable.

\subsection*{Informed Consent.}
Not applicable.

\subsection*{Author Contributions.}
All authors contributed to the study conception, design and implementation. 
Methodology, software, and analysis were performed by Bilal Al-Ahmad, M.~Harshvardhan, and Khaled El-Fakih; conceptualization and supervision by Bilal Al-Ahmad and Khaled El-Fakih; evaluation design and analyses inputs by Anas AlSobeh. 
The first draft was written by M.~Harshvardhan, which was significantly revised by Anas AlSobeh, and all authors reviewed and approved the final manuscript.

\subsection*{Data Availability.}
The ClassEval benchmark is public
(\url{https://github.com/FudanSELab/ClassEval}); study artifacts and the
evaluation code will be made available upon reasonable request.

\subsection*{Code Availability.}
Study artifacts and the evaluation code with results will be made available upon reasonable request.

\clearpage
\newpage

\bibliography{sn-bibliography}

@inproceedings{chen2024chatunitest,
  title={{ChatUniTest}: A framework for {LLM}-based test generation},
  author={Chen, Yinghao and Hu, Zehao and Zhi, Chen and Han, Junxiao and Deng, Shuiguang and Yin, Jianwei},
  booktitle={Companion Proceedings of the 32nd ACM International Conference on the Foundations of Software Engineering},
  pages={572--576},
  year={2024}
}

@inproceedings{pizzorno2025coverup,
  title={{CoverUp}: Effective high coverage test generation for {Python}},
  author={Altmayer Pizzorno, Juan and Berger, Emery D.},
  booktitle={Proceedings of the ACM on Software Engineering (FSE 2025)},
  pages={2897--2919},
  year={2025},
  doi={10.1145/3729398}
}

@article{wang2025projecttest,
  title={{ProjectTest}: A project-level {LLM} unit test generation benchmark and impact of error fixing mechanisms},
  author={Wang, Yibo and Xia, Congying and Zhao, Wenting and Du, Jiangshu and Miao, Chunyu and Deng, Zhongfen and Yu, Philip S. and Xing, Chen},
  journal={arXiv preprint arXiv:2502.06556},
  year={2025}
}

@inproceedings{inozemtseva2014coverage,
  title={Coverage is not strongly correlated with test suite effectiveness},
  author={Inozemtseva, Laura and Holmes, Reid},
  booktitle={Proceedings of the 36th International Conference on Software Engineering (ICSE)},
  pages={435--445},
  year={2014},
  organization={ACM}
}

@article{jia2011analysis,
  title={An analysis and survey of the development of mutation testing},
  author={Jia, Yue and Harman, Mark},
  journal={IEEE Transactions on Software Engineering},
  volume={37},
  number={5},
  pages={649--678},
  year={2011}
}

@inproceedings{guilherme2023initial,
  title={An initial investigation of {ChatGPT} unit test generation capability},
  author={Guilherme, Vitor and Vincenzi, Auri},
  booktitle={Proceedings of the 8th Brazilian Symposium on Systematic and Automated Software Testing (SAST'23)},
  pages={15--24},
  year={2023},
  organization={ACM},
  doi={10.1145/3624032.3624035}
}

@misc{nafis2025llm,
  title={An {LLM}-based framework for automated {Python} test generation and mutation testing evaluation},
  author={Nafis, Sadnan and Walid, Abdullah Al and Anuja, Anuja Sarker},
  howpublished={B.Sc.\ thesis, BRAC University},
  year={2025},
  note={Undergraduate thesis}
}

@article{ouedraogo2024large,
  title={Large-scale, independent and comprehensive study of the power of {LLMs} for test case generation},
  author={Ou{\'e}draogo, Wendk{\^u}uni C. and Kabor{\'e}, Kader and Li, Yewei and Tian, Haoye and Koyuncu, Anil and Klein, Jacques and Lo, David and Bissyand{\'e}, Tegawend{\'e} F.},
  journal={arXiv preprint arXiv:2407.00225},
  year={2024}
}

@article{chudic2026automated,
  title={Automated test suite enhancement using large language models with few-shot prompting},
  author={Chudic, Alex and {\c{C}}al{\i}kl{\i}, G{\"u}l},
  journal={arXiv preprint arXiv:2602.12256},
  year={2026},
  note={To appear, ICPC 2026}
}

@inproceedings{wang2025testeval,
  title={{TestEval}: Benchmarking large language models for test case generation},
  author={Wang, Wenhan and Yang, Chenyuan and Wang, Zhijie and Huang, Yuheng and Chu, Zhaoyang and Song, Da and Zhang, Lingming and Chen, An Ran and Ma, Lei},
  booktitle={Findings of the Association for Computational Linguistics: NAACL 2025},
  pages={3547--3562},
  year={2025},
  doi={10.18653/v1/2025.findings-naacl.197}
}

@article{du2024classeval,
  title={{ClassEval}: A manually-crafted benchmark for evaluating {LLMs} on class-level code generation},
  author={Du, Xueying and Liu, Mingwei and Wang, Kaixin and Wang, Hanlin and Liu, Junwei and Chen, Yixuan and Feng, Jiayi and Sha, Chaofeng and Peng, Xin and Lou, Yiling},
  journal={arXiv preprint arXiv:2308.01861},
  year={2024}
}

@book{halstead,
  title={Elements of Software Science},
  author={Halstead, Maurice H.},
  publisher={Elsevier},
  year={1977}
}

@inproceedings{alves2024detecting,
  title={Detecting test smells in {Python} test code generated by {LLM}: An empirical study with {GitHub} {Copilot}},
  author={Alves, Victor Anthony and Santos, Carlos and Bezerra, Carla and Machado, Ivan},
  booktitle={Proceedings of the XXXVIII Brazilian Symposium on Software Engineering (SBES'24)},
  pages={1--7},
  year={2024},
  organization={SBC},
  doi={10.5753/sbes.2024.3561}
}

@article{schafer2024empirical,
  title={An empirical evaluation of using large language models for automated unit test generation},
  author={Sch{\"a}fer, Max and Nadi, Sarah and Eghbali, Aryaz and Tip, Frank},
  journal={IEEE Transactions on Software Engineering},
  volume={50},
  number={1},
  pages={85--105},
  year={2024},
  doi={10.1109/TSE.2023.3334955}
}

@inproceedings{pan2025aster,
  title={{ASTER}: Natural and multi-language unit test generation with {LLMs}},
  author={Pan, Rangeet and Kim, Myeongsoo and Krishna, Rahul and Pavuluri, Raju and Sinha, Saurabh},
  booktitle={2025 IEEE/ACM 47th International Conference on Software Engineering: Software Engineering in Practice (ICSE-SEIP)},
  pages={413--424},
  year={2025},
  organization={IEEE/ACM}
}

@article{diazarrieta2025comparative,
  title={Comparative analysis of the efficiency of generation of unit test cases: Manual methods versus automation with {LLM}},
  author={D{\'i}az-Arrieta, Ronald H. and D{\'i}az-Monroy, Byron L. and Castillo-Heredia, Luis J.},
  journal={SSRN Electronic Journal},
  year={2025},
  doi={10.2139/ssrn.5185412},
  note={Preprint, not peer reviewed}
}

@inproceedings{yang2025rted,
  title={Reflective unit test generation for precise type error detection with large language models},
  author={Yang, Chen and Wang, Ziqi and Jiang, Yanjie and Yang, Lin and Zheng, Yuteng and Zhou, Jianyi and Chen, Junjie},
  booktitle={2025 40th IEEE/ACM International Conference on Automated Software Engineering (ASE)},
  year={2025},
  note={arXiv:2507.02318}
}

@inproceedings{abdullin2025test,
  title={Test wars: A comparative study of {SBST}, symbolic execution, and {LLM}-based approaches to unit test generation},
  author={Abdullin, Azat and Derakhshanfar, Pouria and Panichella, Annibale},
  booktitle={2025 IEEE Conference on Software Testing, Verification and Validation (ICST)},
  pages={221--232},
  year={2025},
  organization={IEEE}
}

@article{wang2025mutgen,
  title={Mutation-guided unit test generation with a large language model},
  author={Wang, Guancheng and Xu, Qinghua and Briand, Lionel and Liu, Kui},
  journal={arXiv preprint arXiv:2506.02954},
  year={2025}
}

@article{metaach2026,
  title={Mutation-guided {LLM}-based test generation at {Meta} (Automated Compliance Hardening, {ACH})},
  author={Foster, Christopher and Gulati, Abhishek and Harman, Mark and Harper, Inna and Mao, Ke and Ritchey, Jillian and Robert, Herv{\'e} and Sengupta, Shubho},
  journal={arXiv preprint arXiv:2501.12862},
  year={2025}
}

@inproceedings{chattester,
  title={No more manual tests? {E}valuating and improving {ChatGPT} for unit test generation},
  author={Yuan, Zhiqiang and Lou, Yiling and Liu, Mingwei and Ding, Shiji and Wang, Kaixin and Chen, Yixuan and Peng, Xin},
  booktitle={arXiv preprint arXiv:2305.04207},
  year={2023}
}

@inproceedings{siddiq2024using,
  title={Using large language models to generate {JUnit} tests: An empirical study},
  author={Siddiq, Mohammed Latif and Santos, Joanna C. S. and Tanvir, Ridwanul Hasan and Ulfat, Noshin and Al Rifat, Fahmid and Carvalho Lopes, Vinicius},
  booktitle={Proceedings of the 28th International Conference on Evaluation and Assessment in Software Engineering (EASE)},
  pages={313--322},
  year={2024}
}

@inproceedings{gorla2025cubetesterai,
  title={{CubeTesterAI}: Automated {JUnit} test generation using the {LLaMA} model},
  author={Gorla, Daniele and Kumar, Shivam and Roselli Lorenzini, Pietro Nicolaus and Alipourfaz, Alireza},
  booktitle={2025 IEEE Conference on Software Testing, Verification and Validation (ICST)},
  pages={565--576},
  year={2025},
  organization={IEEE}
}

@inproceedings{yang2024evaluation,
  title={On the evaluation of large language models in unit test generation},
  author={Yang, Lin and Yang, Chen and Gao, Shutao and Wang, Weijing and Wang, Bo and Zhu, Qihao and Chu, Xiao and Zhou, Jianyi and Liang, Guangtai and Wang, Qianxiang and others},
  booktitle={Proceedings of the 39th IEEE/ACM International Conference on Automated Software Engineering (ASE)},
  pages={1607--1619},
  year={2024}
}

@misc{humalajoki2024unit,
  title={Unit test generation with {GitHub} {Copilot}, a case study},
  author={Humalajoki, Santeri},
  year={2024},
  note={M.Sc.\ thesis}
}

@misc{bjork2024comprehensibility,
  title={Comprehensibility of editor-integrated {LLM}-generated unit tests},
  author={Bj{\"o}rk, Fredrik and Lindh, Jesper},
  year={2024},
  note={Student thesis}
}

@inproceedings{lops2025system,
  title={A system for automated unit test generation using large language models and assessment of generated test suites},
  author={Lops, Alessandro and Narducci, Fedelucio and Ragone, Azzurra and Trizio, Michelantonio and Bartolini, Claudio},
  booktitle={2025 IEEE International Conference on Software Testing, Verification and Validation Workshops (ICSTW)},
  pages={29--36},
  year={2025},
  organization={IEEE}
}

@article{gao2024automated,
  title={Automated unit test refactoring},
  author={Gao, Yi and Hu, Xing and Yang, Xiaohu and Xia, Xin},
  journal={arXiv preprint arXiv:2409.16739},
  year={2024}
}

@article{ryan2024code,
  title={Code-aware prompting: A study of coverage-guided test generation in regression setting using {LLM}},
  author={Ryan, Gabriel and Jain, Siddhartha and Shang, Mingyue and Wang, Shiqi and Ma, Xiaofei and Ramanathan, Murali Krishna and Ray, Baishakhi},
  journal={Proceedings of the ACM on Software Engineering},
  volume={1},
  number={FSE},
  pages={951--971},
  year={2024}
}

@article{yang2024advancing,
  title={Advancing code coverage: Incorporating program analysis with large language models},
  author={Yang, Chen and Chen, Junjie and Lin, Bin and Wang, Ziqi and Zhou, Jianyi},
  journal={ACM Transactions on Software Engineering and Methodology},
  year={2026},
  doi={10.1145/3748505}
}

@inproceedings{wang2024hits,
  title={{HITS}: High-coverage {LLM}-based unit test generation via method slicing},
  author={Wang, Zejun and Liu, Kaibo and Li, Ge and Jin, Zhi},
  booktitle={Proceedings of the 39th IEEE/ACM International Conference on Automated Software Engineering (ASE)},
  pages={1258--1268},
  year={2024}
}

@article{mutap,
  title={Effective test generation using pre-trained large language models and mutation testing},
  author={Dakhel, Arghavan Moradi and Nikanjam, Amin and Majdinasab, Vahid and Khomh, Foutse and Desmarais, Michel C.},
  journal={Information and Software Technology},
  volume={171},
  pages={107468},
  year={2024}
}

@inproceedings{tackletest,
  title={{TackleTest}: A tool for amplifying test generation via type-based combinatorial coverage},
  author={Tzoref-Brill, Rachel and Sinha, Saurabh and Goldberg, Arie and Pistoia, Marco and Tripp, Omer},
  booktitle={2022 IEEE Conference on Software Testing, Verification and Validation (ICST)},
  year={2022},
  organization={IEEE}
}

@inproceedings{alsobeh2025muae,
  title={{MuAE}: A mutation testing framework for evaluating autoencoders},
  author={Khamaiseh, Samer Y. and Chiacchira, Steven and Alsobeh, Anas and Aljadayah, Abdullah},
  booktitle={2025 IEEE 49th Annual Computers, Software, and Applications Conference (COMPSAC)},
  year={2025},
  organization={IEEE}
}

@article{alazzam2024integration,
  title={Enhancing integration testing efficiency through {AI}-driven combined structural and textual class coupling metric},
  author={Alazzam, Iyad and AlSobeh, Anas M. R. and Melhem, Bushra B.},
  journal={Online Journal of Communication and Media Technologies},
  volume={14},
  number={4},
  pages={e202460},
  year={2024}
}

@inproceedings{alsobeh2025shadowplay,
  title={{ShadowPlay}: Engineering defenses against role-based prompt injection and dependency hallucination in {LLM}-powered development},
  author={AlSobeh, Anas and Gwarzo, Zaid and Shatnawi, Ahmed},
  booktitle={2025 International Conference on Cybersecurity and AI-Based Systems (Cyber-AI)},
  year={2025}
}

@manual{cosmicray,
  title        = {Cosmic Ray: Mutation Testing for Python},
  author       = {{Sixty North AS}},
  year         = {2019},
  note         = {Version latest},
  url          = {https://cosmic-ray.readthedocs.io/en/latest/},
  urldate      = {2026-07-05}
}

@article{al2021jacoco,
  title={Jacoco-coverage based statistical approach for ranking and selecting key classes in object-oriented software},
  author={Al-Ahmad, Bilal and Altaharwa, ISMAIL and Alkhawaldeh, RAMI S and Alazzam, IYAD M and Ghatasheh, NAZEEH},
  journal={Journal of Engineering Science and Technology},
  volume={16},
  number={4},
  pages={3358--3386},
  year={2021}
}

@article{al2018using,
  title={Using Code Coverage Metrics for Improving Software Defect Prediction.},
  author={Al-Ahmad, Bilal},
  journal={J. Softw.},
  volume={13},
  number={12},
  pages={654--674},
  year={2018}
}

\end{document}